\documentclass{svjour3}                     
\smartqed  
\usepackage{natbib}
\bibpunct{(}{)}{;}{a}{}{,} 
\usepackage{graphicx}
\usepackage{hyperref} 
\newcommand{\aap}{A\&A }
\newcommand{\aapr}{A\&A Rev.}

\newcommand{\aj}{Astron. J. }
\newcommand{\apj}{ApJ }
\newcommand{\apjs}{ApJ Suppl. }

\newcommand{\apjl}{ApJ }

\newcommand{\mnras}{MNRAS }
\newcommand{\nat}{Nature }

\newcommand{\pasj}{Publ. Astron. Soc. Japan }

\newcommand{\prd}{Phys. Rev. D}

\begin{document}
\bibliographystyle{plainnat}

\title{Evidence for  the Fifth Element
}
\subtitle{Astrophysical status  of Dark Energy\\ }


\author{Alain Blanchard
}


\institute{A. Blanchard \at
              Laboratoire d'Astrophysique de  Toulouse-Tarbes, UMR 5572, Universit\'e de Toulouse,  CNRS, 14, Avenue E. Belin, F-31400 Toulouse, France\\
              Tel.: +123-45-678910\\
              Fax: +123-45-678910\\
              \email{alain.blanchard@ast.obs-mip.fr}           
}

\date{Received: date / Accepted: date}

\maketitle

\begin{abstract}
Evidence for an accelerated expansion of the universe as it has been revealed ten years ago by the Hubble diagram of distant type Ia supernovae
represents one 
of the major modern revolutions for fundamental physics and cosmology. It is yet unclear whether 
the explanation of the fact that gravity becomes repulsive on large scales should be found within 
general relativity  or within a new theory of gravitation. However, existing evidences for this acceleration
all come from astrophysical observations. 
Before accepting a drastic revision of fundamental physics, it is interesting to critically examine  
the present situation of the astrophysical observations and the possible limitation in their interpretation.  
In this review, the 
main various observational probes are presented as well as the framework to interpret them with
special attention to the complex astrophysics  and theoretical hypotheses that 
 may limit 
actual evidences for the acceleration of the expansion.  Even when scrutinized with sceptical eyes, the evidence for an accelerating universe is robust. Investigation of its very origin appears as the most fascinating challenge of modern physics.

\keywords{Cosmology  \and Dark Energy \and Cosmological models}
\end{abstract}

\section{Introduction}
\label{intro}
Modern cosmology has achieved remarkable progresses during the last fifty years. 
The  general picture originally designed as the "Primeval atom" by Lema\^\i tre and 
which has become the "Big Bang" model according to the word of one of its most famous opponent, F.~Hoyle, is now recognized as the successful
 scientific representation of the world at the large scales (in space and 
in time) we can measure\footnote{It is fair to say that few scientists are still opposed to the "Big Bang" picture.
Most of the serious opponents try to demonstrate that some observational facts,
 most often only one, 
which are coming  in support of the Big Bang may be interpreted in a different 
way and therefore the whole construction has to  be questioned.  It is 
useful to remember that the success of a scientific model is -- in some sense -- measured by the 
number of predictions it leads to and how many are successful. Newton theory 
of gravity is  wrong, but nevertheless it remains a high quality  
scientific theory because of its past (and present) successes. It is in this sense that 
modern cosmology should be regarded as successful, and this will remain in the future, even if it might be regarded as being "wrong"...}.
Construction of this picture has necessitated the successive abandonment  
of philosophical and scientifi ideas, some of which are not only those of physicists but are also 
shared by the more general public. Maybe the first to be given 
up was 
the idea that it is hopeless to try to figure out a global picture of the 
universe.  Although heroic pioneers can be traced back since a long time ago, undoubtedly Einstein is the first one to directly address the question of handling the 
universe as a single physical object with an appropriate tool in hands, general 
relativity (GR).  The fact that one of the very first applications of his theory 
was  Cosmology is an evidence that the cosmological 
question
was central in his thoughts. This is reinforced by the fact that he proposed to 
modify the initial formulation of his theory given the problems he encountered. It is commonly said 
that Einstein introduced the cosmological constant in order to obtain a static 
universe because he was reluctant to the idea of an expanding picture. This formulation 
is  inappropriate: it merely suggests that Einstein was willing to avoid an expanding universe, while he actually wanted to find at least one solution to the cosmological problem with an initial formulation which assumes the universe to be stationary, there is no indication in his seminal paper that he wished to 
reject a non-stationnary universe:   he actually started his paper 
by  discussing   the 
problem of having  mass at large distances in the Newtonian approach, 
noticing that this leads to divergence of the  potential. He insisted that 
this  would lead to unacceptably large velocities for stars. He also quoted that 
this can be cured by assuming a correcting term to the Newtonian potential  equation:
\begin{equation}
\nabla^2 {\varphi} - \lambda {\varphi} = 4\pi G \rho
\label{}
\end{equation}
and then proposed to modify his initial theory with the addition of the cosmological term $\Lambda$. This allowed him to
 construct the first 
relativistic cosmological model, the Einstein solution, which is spatially 
closed (because being spherical) and static. In 1919, de Sitter discovered a new
 solution to Einstein equation which was written in a stationary form\footnote{The choice of the coordinates system lead  to a form of the metric for which the coefficients are constant.} and 
contains no matter (but a non-zero cosmological constant). It is only a few 
years after, in 1925, that Lema\^\i tre identified the de Sitter solution  
to  an homogeneous  expanding universe \citep{Lemaitre1925}.
Friedmann \cite{Friedmann1922,Friedmann1924} found the  general homogeneous solutions, providing the equation 
for the scale factor $R(t)$ and recognised their  expanding nature. It is 
somewhat surprising that his work has remained totally unnoticed, despite a 
controversy with Einstein. During this period  it is clear that 
the nature of the redshift of what Hubble had identified as extragalactic 
nebulae became a question addressed by many astronomers. Slipher's 
discovery was probably much more intriguing now that the nature of nebulae 
had been identified. Eddington is often mentioned as  the first astronomer
to have noticed a possible connection between redshift and the de Sitter 
solution \cite{Lemaitre1925}.  Carl Wirtz was clearly looking for a relation between redshift 
and distance \cite{Wirtz1} and had in mind the possible cosmological information it might provide \cite{Wirtz2}. Lema\^\i tre reestablished the equations Friedmann derived, 
showing 
that expanding solutions were leading to a redshift proportional to the distance.
He proposed that this effect  was the origin of known redshift and provided the first estimation of the Hubble constant \cite{Lemaitre1927}. In 1929,  Robertson published the now so called Robertson-Walker metric. In the same paper he mentioned that  distant sources appear 
to have a frequency shift.

\section{Basics of  Friedmann-Lema\^\i tre Models}
\label{sec:2}
The fundamental idea of the geometrical theory of gravity starts from the 
fact that we can assign four coordinates to any event observed in our 
vicinity, for instance in Cartesian coordinates~: $(x,y,z,t)$. Locally, 
space appears flat to be. However this does not determine of the 
geometry  of space at larger scales~: local observations put us in the 
same situation 
that led people to think the earth was flat~: the fact that we can describe our vicinity by a flat map does not determine the actual geometry on larger scales. Let us take the 
line element of a homogeneous 3D space which can be shown\footnote{It is an instructive exercise to start from an Euclidean 4D space $x,y,z,u$ and derive the line elements $d l^2$ on the 3D sphere ($x^2+y^2+z^2+u^2 = R^2$) in internal spherical coordinates ($\tilde{r} = \sqrt{x^2+y^2+z^2},\theta,\phi$).} to be :
\begin{equation}
d l^2 = \tilde{r}^2 (d \theta^2 + \sin ^2 \theta d \phi^2) + {{d \tilde{r}^2}\over
{1-k\left({\tilde{r}}\over {R}\right)^2}}
\label{}
\end{equation}
where $k$ is $-1,0,1$ according to whether space is hyperbolic, 
flat or spherical.
$R$ is a characteristic size (in the spherical case, that is the
radius of the 3D-sphere embedded in a 4D space).

We then add the time as the fourth coordinate to build the equivalent 
of the Minkowski space-time element of special relativity and get the Robertson-Walker 
(RW) line element after the change of variables
${{\tilde{r}}\over {R}} \rightarrow r$:
\begin{equation}
d s^2 = - c^2d t^2 +R(t)^2[r^2(d\theta^2+\sin^2\theta d\phi^2)
+ {{d r^2}\over {1-kr^2}}] 
\label{}
\end{equation}

\subsection{Topology}

The above line element describes  the local shape of space~: the 
curvature 
is only a local property of space, 
 but does not tell us about the {\it global} shape of space. 
For instance, the Euclidean plane is an infinite flat surface while 
the surface of a cylinder is a 2D-space  which is flat everywhere but 
is finite in one direction.  GR will in principle allow us to  derive
the local geometry of space and its dynamics, but does not 
specify of the global topology of space. Only direct observations 
would allow  to test what the topology actually is. Of course this 
will not be possible on scales much larger than what can be observed 
(the horizon). We 
can therefore hope to prove that the Universe is finite, if it is 
small enough, but we could not know whether we are
in a finite Universe of which the scale is larger than the horizon, 
or whether we are in an infinite Universe. The interest in the topic of the cosmic topology, with possible observational signature,  has been recently revived \cite {ll,luminet}

\subsection{Dynamics}

The function $R(t)$ which appears in the RW line element, is totally 
independent of any further geometrical consideration. It can be specified 
only within a theory of gravity. 
 The basic 
equation of GR relates the geometrical tensor $G_{ij}$ to the 
energy-momentum tensor $T_{ij}$
\begin{equation}
G_{ij} = R_{ij}
-{{1}\over {2}} g_{ij}R = 8\pi G T_{ij}
\label{eq:Einstein}
\end{equation}
where  $g_{ij}$ is the metric tensor, $R_{ij}$ is the Ricci tensor, $R$ the Ricci scalar. For a perfect fluid, there exists a coordinates system, called the comoving coordinates, 
in which the matter is at rest and the tensor $T_{ij}$ is diagonal 
with $T_{00}= \rho$ and $T_{11}=T_{22}=T_{33}= p$,
$\rho$ being the density and $p$ the pressure. A fundamental aspect 
of GR is that the source of gravity includes explicitly a term coming 
from the pressure~: $\rho + 3 p/c^2$. Finally, there is an analog of 
the Gauss theorem, that is the Birkhoff's theorem \cite{Birkhoff}\footnote{Apparently, this theorem should be named Birkhoff-Jebsen, as it has been published two years earlier by an  Norwegian physicist, J.T. Jebsen\cite{BirkhoffR}.}: if the matter 
distribution is spherical then the evolution of the radius of a given 
shell of matter depends only on its content.

 From the above rules, we can easily derive the equation for $R(t)$.
Let us consider a spherical region of radius $a$ in a  homogeneous
distribution of matter. The equivalent Newtonian acceleration is~:
\begin{equation}
{d ^2 a \over d t^2} = g
\label{}
\end{equation}
with the acceleration being generated by the ``mass''  $M(a)$ of the above 
spherical region~:
\begin{equation}
g = -{GM(a)\over a^2} = - {4\over 3}\pi G(\rho+3p/c^2)a
\label{}
\end{equation}
The density term includes the effect of kinetic energy ($E = mc^2$!). 
Writing total energy ($E_t$) conservation  inside the volume of the 
sphere from elementary thermodynamics gives~:
\begin{equation}
d (E_t) = d (\rho Vc^2) = -p d V
\label{}
\end{equation}
leading to~:
\begin{equation}
\dot{\rho} = -3({p\over c^2} + \rho){{\dot{a}}\over {a}}
\label{eq:en}
\end{equation}
 From these two equations, the pressure can be eliminated, and, after 
having multiply both terms by $\dot{a}$, the differential  equation 
can be easily integrated. This leads to the following equation~:
\begin{equation}
\left({\dot{a} \over a}\right)^2
= {{8\pi G\rho}\over {3}} - {{Kc^2}\over {a^2(t)} }
\label{}
\end{equation}
The last term corresponds to the constant of integration. Its value 
cannot be specified, depending on the initial conditions.
The form of the above equation is independent of the radius $a$ of the sphere and 
the solution $a(t)$ should be proportional to  the 
quantity $R(t)$. $R(t)$ should also be solution of an equation of the same form,  the constant $K$, which depends on the radius $a_0$, being related to the constant $k$ which is 
involved in the Robertson-Walker metric element, something which can be established only within GR~:
\begin{equation}
\left({\dot{R} \over R}\right)^2
= {{8\pi G\rho}\over {3}} - {{kc^2}\over {R^2(t)} }
\label{}
\end{equation}
This relation  expresses  the
link within the framework of GR between the geometry and the material content of the Universe.
In order to specify completely the function $R(t)$, one needs an 
equation of state for the content of the Universe. The two cases 
often seen in cosmology are the dust case ($p = 0$) and the radiation 
dominated regime ($p = {{1}\over {3}}\rho c^2$).

\subsection{Vacuum and the Cosmological Constant}

Vacuum is a particular medium, and one should wonder what is the 
 equation of state of this medium. Naively, one would think that the 
equation will be  $p_{\rm v}=0$ and $\rho_{\rm v} = 0$. However, let us try to derive 
the vacuum equation of state from first principles. As in
classical thermodynamics let's assume that we have a piston with 
vacuum in it. We also assume that simple vacuum ($p=0,\rho=0$) is 
present outside.

 The energy inside the piston is $E = \rho_{\rm 
v}c^2 V$. If the volume changes by a small amount the net energy 
change is:
\begin{equation}
d E = d (\rho_{\rm v}Vc^2) = \rho_{\rm v}c^2d V
\label{}
\end{equation}
this change is equal to the work of the pressure:
\begin{equation}
d E = -p_{\rm v} d V
\label{}
\end{equation}
so the equation of state is:
\begin{equation}
p_{\rm v} = - \rho_{\rm v} c^2
\label{}
\end{equation}
As one can see, the conditions $p\geq 0$ and $\rho \geq 0$  ensure that 
the simple solution is the only  one. However, there is nothing which 
imposes these conditions for the vacuum, and we can therefore decide 
to keep such a possible term. This can be directly translate in the 
equations governing $R(t)$   by introducing the following 
constant~:
\begin{equation}
\Lambda = 8 \pi G \rho_{\rm v}
\label{}
\end{equation}
Such a term is called the cosmological constant and has been 
historically introduced by Einstein as a modification of his 
original theory. It appears as an additional term in the left hand side of eq. \ref{eq:Einstein}. We have  recovered the two usual
Einstein-Friedmann-Lema\^{\i}tre (EFL) equations:
\begin{equation}
\left({{\dot{R}}\over {R}}\right)^2 = {{8\pi G\rho}\over {3}} -
{{kc^2}\over {R^2} }+ {{\Lambda}\over {3}} 
\label{eq:EFL1}
\end{equation}
and
\begin{equation}
{{\ddot{R}}\over {R}} = -{{4\pi G}\over {3}}\left(\rho+\frac{3p}{c^2}\right) + {{\Lambda}\over {3}} 
\label{eq:EFL2}
\end{equation}
There are several ways to write the EFL equations. For instance we used energy conservation, eq. \ref{eq:en}, which can be kept instead of one of the above eqations. It can  also be useful to use:
\begin{equation}
{{\dot{H}}} = -{{4\pi G}}\left(\rho+\frac{p}{c^2}\right) +{{kc^2}\over {R^2} }
\label{eq:EFL3}
\end{equation}
Only two independent equations were obtaineded, while three unknown quantities are to be determined ($R(t), \rho(t), p(t)$), therefore we need a further constraint which is provided by the equation of state 
$F(\rho,p)=0$.

 It is quite usual to write the fundamental cosmological quantities as non-dimensional quantities which depend on redshift. The following notations are very common~:
\par $H = {\dot{R}\over R}$, the Hubble parameter,
\par$\Omega_M = \Omega = {8\pi G\rho\over 3H^2}$ the density parameter,
\par$q = -{\ddot{R}R\over \dot{R}^2}$, the deceleration parameter,
\par$\Omega_v= \Omega_\lambda = \lambda  ={ \Lambda\over 3H^2}$, the (reduced) 
cosmological constant,
\par  $\Omega_c = - {kc^2\over H^2 R^2} = - \alpha $, the curvature parameter.\\

  Quantities are labeled with a $0$ when they are referred to their 
present value. For instance the  present day value of the density 
parameter $\Omega$ is $\Omega_0$.
With the above notations, the first EFL equation \ref{eq:EFL1} then reads~:
\begin{equation}
\Omega_{M}+\Omega_{c} + \Omega_{\lambda } = 1
\label{}
\end{equation}
or:
\begin{equation}
\alpha = \Omega + \lambda - 1
\label{}
\end{equation}
so that the "radius of the Universe" can be written  :
\begin{equation}
R = {c \over H} {1 \over \sqrt{|\alpha}|}
\label{}
\end{equation}
while the Hubble constant evolution is given by:
\begin{equation}
H^2 = H_0^2(\Omega_{M}(1+z)^3+\Omega_{c}(1+z)^2 + \Omega_{\lambda })
\label{H(z)}
\end{equation}

\subsection{Important quantities needed for observations}

In this section we only need to work in the framework of a geometrical
theory of space-time, in which the trajectories of light rays are assumed to
be the null geodesics. Let us have a comoving spherical coordinate system
$(r,\theta,\phi,t)$  the observer being at the origin of the spatial coordinates
$(r=0,\theta=0,\phi=0, t_0)$, let assume that the observed source
is emitting light at the coordinates $(r_{\rm S},\theta=0,\phi=0,t_{\rm S})$,
and let $r(t)$ be the trajectory of the emitted  photons. As this trajectory
is a null geodesic, we have:
\begin{equation}
c^2d t^2 - R^2(t){d r^2\over 1-kr^2} = 0
\label{}
\end{equation}
so the variables can be separated and the integration over $r$ is analytical:
\begin{equation}
\int_{t_{\rm S}}^{t_0}{c dt\over R(t)} =
\int_0^{r_{\rm S}}{d r\over \left(1-kr^2\right)^{1/2}}
= S_{k}^{-1}(r_{\rm S})
\label{}
\end{equation}
with:
\begin{equation}
S_{k}(r_S)  =  \cases{~\sin (r_{\rm S}) &{\rm if } {\em k} = +1\cr
                       r_{\rm S} &{\rm if } {\em k} = 0 \cr
                       \sinh (r_{\rm S}) &{\rm if } {\em k} = -1\cr}
\label{eq:Sk}
\end{equation}

When the distance is small with respect to  $R_0$ we just have $S_{k}^{-1}(r) \sim  r$.
\subsection{The Redshift}

In order to derive the observed frequency $\nu_0$ of the light 
from a source emitted at the frequency $\nu$, we consider the 
trajectory of a second light ray emitted at the time $t_{\rm S}+{1\over 
\nu}$. As the source is
comoving its coordinate is unchanged and we have:
\begin{equation}
S_{k}^{-1}(r_{\rm S}) = \int_{t_{\rm S}}^{t_0}{c d t\over R(t)}
= \int_{t_{\rm S}+1/\nu}^{t_0+1/\nu_0}{c d t\over R(t)}
\label{eq:r}
\end{equation}
which implies:
\begin{equation}
{\nu_0\over \nu}={\lambda_{\rm S}\over \lambda_0}={R_{\rm S}\over R_0}
={1\over 1+z}
\label{eq:nu}
\end{equation}
where $z$ is the redshift.
This is the standard formula for the cosmological shift of the frequencies.
This result shows that the redshift $z$ is a natural consequence of the 
expansion.

\subsection{The proper distance}

In GR,  space changes with  time, and there is no proper time, so that the ``intuitive'' notion of 
distance between two points is not a well defined quantity. 
Therefore the various 
methods to measure the distance between  an observer and  a given  source 
 give 
different answers. 
  The proper distance -- between the source and the observer -- can be seen as a
distance measured by a set of rulers at time $t$. The distance element is given by :
\begin{equation}
d l^2 = d s^2 = R(t)^2 { d r^2\over 1-kr^2}
\label{}
\end{equation}
so that the proper distance is~:
\begin{equation}
D_p = R(t) S_{k}^{-1}(r_{\rm {S}})
\label{}
\end{equation}
The fact that this distance changes with time is the direct
consequence of the expansion of the Universe. We can now examine how this length changes 
with time :
\begin{equation}
\dot{ D_p} = \dot{R}S_k^{-1}(r) 
\label{}
\end{equation}
so that the source is {\it actually receding} from the observer with a speed:
\begin{equation}
V = \dot{ D_p} = {\dot{R}\over R} D_p = H D_p
\label{}
\end{equation}
The fact that this speed could be larger than the speed of light should not be
considered as a problem: this speed can be measured but cannot transport
  information faster than light. When the distance is small, the Doppler 
frequency shift is :
\begin{equation}
{\delta \nu\over \nu}= {\dot{R}\over R}\delta t = H {D\over c} = {V\over c}
\label{}
\end{equation}
so that the shift is the one corresponding to the Doppler 
shift associated with the above velocity. For large distances, the total shift results from the product of small Doppler shifts and  the redshift  is therefore 
purely kinematic. The physical nature of the expansion has been recently the subject of interesting discussions \cite{Chodo2007,abra,Peacock2008,cook}.

\paragraph{Comoving  distances}

It is sometimes useful to refer to comoving distances\footnote{This could also be confusing! }. The
comoving  distance $D^c(z)$ associated to the distance $D(z)$ is :
 \begin{equation}
D^{\rm c}(z) = \frac{R_0}{R} D(z) = (1+z) D(z)
\label{}
\end{equation}
In the case of the proper distance, this becomes:
 \begin{equation}
D^{\rm c}_p(z) = R_0 S_{k}^{-1}(r) = \int_{t_{\rm S}}^{t_0}{c d t\over R(t)/R_0} =  c\int_{0}^{z}\frac{dz}{H(z)}
\label{}
\end{equation}

\subsection{The angular distance}

Let us suppose that we observe a ruler orthogonal to the line of sight.
The extremities of the ruler have the coordinates $(r,0,0,t_{\rm {S}})$ and
$(r,\theta,0,t_{\rm {S}})$. The proper length $l$ between the extremities is:
\begin{equation}
l^2 = d s^2 = R(t_{\rm {S}})^2r^2\theta^2
\label{}
\end{equation}
which provides  the relation between the angle $\theta$ and the
length $l$ and thereby the angular distance defined by:
\begin{equation}
D_{\rm ang} =  \frac{l}{\theta}= R(t_{\rm {S}})r
\label{}
\end{equation}

\subsection{The luminosity distance}

Let us assume that we observe a source with an absolute luminosity
$L$ through a telescope with a diameter $d$ and let us
choose a coordinates system which is centered on the source.
  Let $\theta$ be the angle between two rays reaching two points
 diametrically opposite  
on the telescope. We have $d = R(t_0)r\theta$.
  The energy emitted by the source that reaches the telescope is~:
\begin{equation}
s = {L \over4\pi} \times {\pi \theta^2 \over 4 }
\label{}
\end{equation}
When observed, the energy of  photons has been shifted by $1/(1+z)$ but also
the frequency at which they arrive is reduced by the same factor.  Therefore
the flux (energy per unit time and unit surface) one gets is:
\begin{equation}
f = {s\over \pi l^2/4 }{1\over (1+z)^2} = {L\over 4\pi R(t_0)^2r^2(1+z)^2} =
{L\over 4\pi D_{\rm lum}^2}
\label{}
\end{equation}
This relation provides the luminosity distance:
\begin{equation}
D_{\rm lum} = R(t_{0})r(1+z) = R(t_{\rm S})r(1+z)^2 = D_{\rm ang}(1+z)^2
\label{}
\end{equation}
\subsection{Distance along the line of sight}

We consider here the length along the path of a photon trajectory. The length element is
\begin{equation}
dl = c dt= c{dR\over \dot{R}}= - {c\over H(z)}{d z\over 1+z}
\label{eq:dl}
\end{equation}
This relation is useful to write the volume element.

\subsection{The age of the universe}

The general expression of time interval is :
\begin{equation}
dt= {dR\over \dot{R}}= - {1\over H(z)}{d z\over 1+z}
\label{eq:dt}
\end{equation}

\section{Some solutions of the  EFL equations~: relativistic cosmological models}

The various possible theories of gravitation provide different functions $R(t)$
and through the above tests may in principle be distinguished by
observations. However, it is easy to check that the difference only occurs
at high redshift. In practice these tests may not be discriminant
because the observations of distant objects are difficult and because
the universe at high redshift is younger, so any object at high redshift 
is likely to be different from  to-day because of evolution.  In other words, 
general  geometrical tests rely on the assumption that the typical evolution time scale 
of the objects undr study is much  {\em larger}  than the age of the universe.

One must also underline that as already mentioned the EFL equations can be solved only once the equation of state is specified, i.e. a relation  
between $p$ and  $\rho$ is adopted. It is now quite common to specify this relation 
by the $w$ parameter:
\begin{equation}
p = w \rho c^2
\label{}
\end{equation}
If $w$ is constant, then  eq. \ref{eq:en} allows to find the evolution of the
field density:
\begin{equation}
\rho(z) = \rho_0(1+z)^{3(1+w)}
\label{}
\end{equation}
 while for a general function $w(z)$ one has:
\begin{equation}
\rho(z) = \rho_0\exp \left( \int_0^{z}{3\left(1+w\left(z\right)\right)}\frac{dz}{1+z}\right)
\label{}
\end{equation}

There are three particular regimes :  the matter dominated one,  the 
radiation dominated one and the vacuum dominated one.

In the matter dominated case one has $p =0$, so $w =0$ and  the mass (per comoving volume)  is conserved : 
\begin{equation}
 \rho 
a^3 = \rho(1+z)^{-3} =  \rm cste
\label{}
\end{equation}
while in the pressure dominated case  $p =\frac{1}{3}\rho c^2$, so $w =\frac{1}{3}$
\begin{equation}
\rho a^4 = \rho(1+z)^{-4} =  \rm cste
\label{}
\end{equation}
Finally in the vacuum dominated case  $p =-\rho c^2$, so $w = -1$ and
\begin{equation}
\rho   =  \rm cste
\label{}
\end{equation}
In some models $w < -1$, and therefore the density of the universe increases when it expands. 

\subsection{Case $(\lambda_0 = 0, p=0)$}

When the cosmological constant is zero, there are three types of solutions:

a) when the density is above the critical density:
\begin{equation}
\rho > \rho_c =  {3H_0^2\over 8\pi G} = 2.\ 10^{-29} h^2 \ \rm g/cm^3
\label{}
\end{equation}
the spatial solution is the spherical space. The function $R(t)$ grows from
zero to a maximum value then a collapse phase follows to zero.

b) when the density is equal to the critical density, the solution is 
named the Einstein-de Sitter universe and $R(t)$ is
simple~:
\begin{equation}
R(t) = R_0\left({3\over 2} H_0t\right)^{2/3}
      = R_0\left({t\over t_0}\right)^{2/3}
\label{}
\end{equation}
with $t_0 = {2\over 3}H_0^{-1} = 1/(6\pi G \rho_c)^{1/2}$

c) when the density is below the critical density, the function $R(t)$ grows
from zero to infinity.
It is easy to check from the EFL equations
that the function  $R(t)$ behaves like $t$ when $R$ is large.\\

The behavior of $R(t)$ can be found when $t\rightarrow 
0$ independently
of the model: $R(t)\propto t^{2/3}$.

Finally, the relation between the comoving coordinate $r$ and the redshift
can be expressed:
$$
R_0r = {c\over H_0}{2\over \Omega_0^2}
{\Omega_0(1+z)+2-2\Omega_0-(2-\Omega_0)\sqrt{1+\Omega_0z}\over 1+z}
$$
This is known as the Mattig relation. Others useful quantities can 
be  found in Weinberg recent text book \cite{Weinberg2008}. 

\subsection{Case $(\lambda_0 > 0, p = 0)$}

\begin{figure}
\begin{center}
\includegraphics[angle=90,width=1.0\textwidth]{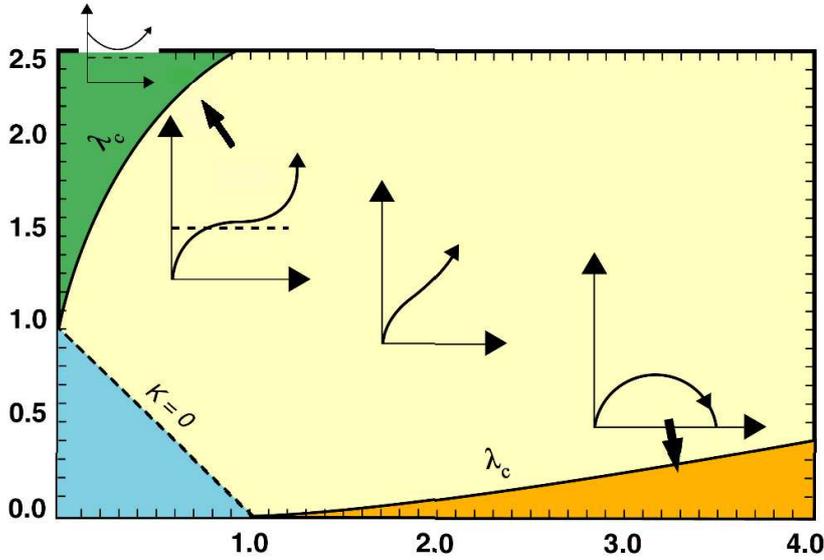}
\caption{ Evolution of the scale-factor $R(t)$ with  time, in FL models according to the  values of the parameters $\Omega_M$ and  
 $\Omega_\Lambda$. One notes that above the critical line universes has  
 no initial singularity; on the right side of the second branch of the  critical line 
are the re-collapsing models. \label{fig:models}
}
\end{center}
\end{figure}

There are many possibilities when a cosmological constant is allowed.
To specify a cosmological model, it is customary to specify two "observables":
$\Omega_0$ and $\Omega_\lambda $. For instance, the cosmological  view of the Friedmann-Lema\^ \i tre 
models is summarized in figure \ref{fig:models}.

The look back time, i.e. the time
since the epoch corresponding to the redshift $z$ con be obtained from \ref{eq:dt}:
\begin{equation}
\tau (z) = {1\over H_0}\int_{1}^{1+z} {du \over u 
\sqrt{\Omega_0u^3 -\alpha_0u^2 + \lambda_0}}
\label{eq:tau}
\end{equation}
where $u = 1+z$. When $\lambda \leq 0$ it can be shown from the 
expression of $\ddot R$ that
\begin{equation}
t_0 = \tau (\infty) \leq {1\over H_0}
\label{}
\end{equation}

\subsection{Radiation dominated case}

As we have seen the density associated with radiation evolves
according to $ \rho_{\gamma} a^4 = \rm cste$.
It is clear that this radiation term will be dominant over the matter
term for small $a$, that is
at the ``beginning''. In this regime, if we neglect other terms in the
EFL equation, we have:
\begin{equation}
\dot{R} = \left({8\pi G\rho_1\over 3}\right)^{1/2}{R_1^2\over R}
\label{}
\end{equation}
so that $R\propto t^{1/2}$

\subsection{Basics of structure formation}

Structure formation in an expanding Universe through gravtational instability 
can be traced back to the early 
works of Lema\^{\i}tre \cite{Lemaitre1933}, who established the spherical 
solution and discussed its relevance to cluster formation. In modern 
cosmology, linear and non linear regime ($\displaystyle {\delta \rho} {\rho} \gg 1$) are used to obtain useful constraints.

\paragraph{The gravitational instability picture} 
 
The linear regime of small perturbations (both in size and amplitude) in the matter field  ($\displaystyle \rho(x) = \rho \left(1+\delta(x)\right)$) in the absence of pressure can be derived from newtonian equations \cite{Peebles1980}:
\begin{equation}
\ddot{\delta}+2H\dot{\delta}= 4\pi G\rho_m \delta = \frac{3}{2}\Omega_m H^2 \delta
\label{eq:growthrate}
\end{equation}
Because there is no spatial derivates in the equation, the evolution with time 
is independent of the scale (this does not hold any more when the pressure is 
 taken into account). 
There are generally two distinct solutions to this equation, one is a decaying mode and the second one is a growing mode which is of relevance for cosmology. The growing mode is usually written as $D(t)$ with a normalization such that the present day value is one: $D(t_0) = 1$. The evolution of a perturbation can then be written:
\begin{equation}
 \delta (t) = D(t)  \delta_0
\end{equation}
where $\delta_0$ is the linear  amplitude that the fluctuation would have 
today.  The qualitative behavior of solutions of eq. \ref{eq:growthrate} is easy to understand: if  $\Omega_m \ll 1 $  the right hand side is  zero and the growing mode is frozen, while if $\Omega_m \sim 1 $ there is a growing mode with $D(t) \propto a(t)$. The form of eq. \ref{eq:growthrate} implies that once the expansion rate history $(H(z)$ and the present day density of the universe $\rho_0$ are known, the solution $D(t)$ is entirely determined (the matter density parameter can be inferred from its definition $\displaystyle\Omega_m(z)= \frac{8\pi G\rho_m}{3H^2}= \Omega_0 \frac{(1+z)^3}{H(z)^2}$).
  Therefore having a measure of the growing rate of perturbations is a direct test of the relevance of Newtonian gravity on the scale under consideration. Theories beyond GR might lead to growing rates that 
depend on the scale and departs from their standard values.

\paragraph{Non baryonic dark matter and  Cold Dark Matter scenario}

In the early universe the growth rate of fluctuations is more complex : 
pressure has to be taken into account, and there are at least two matter 
contents to take into account: baryons and the non-baryonic matter. These ingredients will introduce  scales in the dynamics of fluctuations. Ultimately these scales will leave visible imprints in observables quantities~: the fluctuations of the cosmic background radiation and the power spectrum of the matter distribution (or equivalently in the correlation function). The first scale comes from the difference in the dynamics for fluctuations larger than the horizon compared to the ones that are smaller. Indeed, when photons (and other relativistic particles)  dominated the density of the universe the right hand side  of eq. \ref{eq:growthrate} 
is negligible, and therefore the growing mode is frozen. However,  on  scales larger than the horizon the dynamics should be evaluated within a full GR treatment. Intuitively, on these large scales the radiation acts as a source of gravity which will allow the fluctuations density 
to grow.  Therefore a scale appears corresponding to the scale of the horizon.
When the universe enters the matter dominated area, fluctuations on all scales
 will grow provided that the pressure term is negligible, i.e. for the non-baryonic matter. The scale which is imprinted is therefore the size of the horizon at the time of equality between matter and radiation and is characteristic of Cold Dark Matter (CDM). However,  the baryons are still tightly coupled to the photons and thereby their dynamics is under the control of the pressure term to be added  in the rigt hand side of eq. \ref{eq:growthrate}. This coupling stops when the matter becomes neutral enough ($z\approx 1500$). A second scale is therefore imprinted in the baryonic component corresponding to the scale of the horizon of sound's velocity at the epoch of recombination. Although the baryonic matter is subdominant, the distribution of dark matter will nevertheless reflect this scale. This feature appears as oscillations in the power spectrum, but more distinctly as a peak in the correlation function \cite{Matsubara2004}.  In 
practice, the dark mater distribution is described by its linear 
power spectrum $P_{DM}(k)$ at present epoch corresponding to the initial power spectrum $P_{i}(k)$ modified by the evolution. In the linear regime  the various 
wave numbers are independent so:
\begin{equation} 
P_{DM}(k) =  P_{i}(k) T_{DM}(k)
\end{equation} 
The nature of the initial fluctuations, the nature of dark matter (hot, warm, ...) are other 
ingredients that determine the  shape of the transfer function. 
 In practice, the computation of the transfer function is to be done numerically. 
In the cold dark matter case (CDM) analytical functional forms have been provided
\cite{PeeblesCDM,Bardeen1986,EisensteinHu1998} including the detailed baryonic features \cite{EisensteinHu1998}.

\section{The Quest for cosmological parameters}
\label{sec:1}


\subsection{ Hubble diagram }

 Looking for a possible relation between apparent velocities and 
distances was a scientific question that few astronomers addressed before 
Hubble's discovery. 
The determination of Hubble's constant have been the subject of considerable effort over the next eighty years following his discovery. One of the major scientific question that the Hubble Space Telescope was aimed to address was precisely 
the determination of the distance of close galaxies and thereby to offer an accurate measurement of the Hubble constant. The final value obtained from the HST key project to measure the Hubble constant concluded $H_0 = 72 \pm 8$ km/s/Mpc \cite{Freedman2001}, also the debate is not entirely closed as Sandage and Tammann still concluded to a lower value: $H_0 = 62.3 \pm 4$ km/s/Mpc \cite{Tammann2008} although these values are marginally consistent given the quoted  --~systematic~-- uncertainties\footnote{The Hubble constant is traditionally noticed: $H_0 = h 100$ km/s/Mpc.}.

\subsection{ Going further and other geometrical tests}

The first cosmological tests which have been proposed were based on the relation 
between some observable quantity  to a corresponding intrinsic property of the source 
(like the relation between the apparent luminosity and the intrinsic luminosity). These tests are essentially geometrical in nature, they  involve the coordinate $r$ given by eq. \ref{eq:r} or the 
relation between the coordinate $r$ and the redshift $z$:
 \begin{equation}
r  =   S_k\left(\frac{c}{R_0}\int_{0}^{z}\frac{dz}{H(z)}\right)
\label{}
\end{equation}
the dependence on cosmological parameters coming from the difference in the 
expansion rate according to eq. \ref{eq:EFL1}. The Hubble diagram which has been extended to high redshift 
thanks to type Ia supernovae is the most popularized example of these geometrical tests.
It is also rather intuitive that the Hubble diagram which expressed the speed 
of the expansion will provide direct information on the rate of 
acceleration/deceleration when extended to objects distant enough to be  
seen at  
appreciably early epoch of the history of the universe. A reliable extension of the Hubble 
diagram to high redshift has been made possible thanks to the use of type Ia 
Supernovae. SNIa could have a maximum luminosity ($M\sim -19.5$) comparable to 
that of an entire galaxy. Furthermore there is a relation between the decline 
rate and the intrinsic luminosity  making them suitable for distance 
measurements
at cosmological scale. Because SNIa are rare, large sky area have to be surveyed 
on a regular basis to collect samples of  SNIa. At the end of last century, 
two groups have independently investigated
  the distant SNIa Hubble diagram and 
concluded that supernovae at redshift $\sim 0.5$ were dimmer by $\sim$ 0.2 mag 
compared to what was expected in a unaccelerated universe. This was interpreted as an evidence for an accelerated expansion\footnote{There is an intense debate on the ``discovery'' of 
acceleration from SNIa Hubble diagram in order to specify who did what and 
who said what. The history of the research program developed within the SCP 
is summarized by Gerson Goldhaber\cite{SCP3}. An early popular scientific report 
on SCP talk at AAS meeting January 1998 meeting has been published by James  
Glanz (Science, 279, 651), while some  view from the High Z team  by
 Robert Kirshner \cite{Kirshner2002} is available at \url{http://www.cfa.harvard.edu/~rkirshner/whowhatwhen/Thoughts.htm}. It is also interesting to mention that claim for evidence of an accelerated expansion from a totally different technique has been presented at the same AAS meeting: \url{http://www.bk.psu.edu/faculty/daly/PU98.pdf}. These data taken at face value were pointing to a low density universe, but only the flatness of the universe as already evidenced by CMB fluctuations would have allowed to conclude on the actual acceleration of the expansion.  
} \cite{Riess98,SCP2,SCP1,HZT2}.


\subsection{The age of the Universe}

Formally, we have no direct information on the actual age of the universe $t_0$.
But there are several astrophysical objects for which 
an age can be derived. The most common one is probably  the age of globular clusters. There are different limitations which make determinations of absolute ages problematic. Distance estimations is for instance one limitation. 
Hipparcos results have helped, but the uncertainty remains significant.
Recent estimate for  the age of globular clusters was  given to be 
$12,6^{-2.2}_{+3.4} (95\%)$ Gyr \cite{KrausChaboyer}, an age consistent with that of  the star CS 31082-001
estimated to be $14. \pm 2.4$ Gyr, based on the decay of U-238 \cite{Hill2002}.
It has been noticed since the beginning of cosmology that age estimations 
were high given the preferred value  of the Hubble constant. This has been noticed by Lema\^\i tre when debating with Einstein, 
Peebles \cite{Peebles93} and many others since, providing thereby an argument in favor of the cosmological constant which holds until the late 90's \cite{KraussTurner,Ellis2002}. However, this argument is relatively weak because of the difficulty to
achieve reliable astrophysical estimates of $H_0$. With modern best values, the product $H_0t_0$ can be evaluated numerically:
\begin{equation}
H_0t_0 \approx 0.102ht_{\rm Gyr} 
\end{equation}
with $h =0.72$ one gets $H_0t_0 \sim 0.93$  while only values greater than 1 
would  request an accelerated universe. Although, it is important to have a 
consistency check, improvement by one order of magnitude in precisoin will be necessary in 
order that age and Hubble constant values could be combined to 
provide  constraints on cosmological parameters competitive with current constraints.

\subsection{The Cosmic Microwave Radiation}
 \paragraph{Spectrum and uniformity}
The  measurement of the CMB spectrum to check for the black-body 
shape of this spectrum  was one of the most important cosmological test 
awaited for since the discovery of the background radiation. Two 
high quality measurements have been published in 1990 \cite{Gush1990,Firas}, 
while the FIRAS final results were remarkably accurate \cite{Fixsen}: any departure from a blackbody should represent only a tiny perturbation of the total energy: $\delta E/E \leq 10^{-4}$. 

An interesting  application of cosmic background
radiation is to provide a test of the reality of the expansion: if we are
able to measure the temperature of the background at higher redshift
it should scale according to:
\begin{equation}
T(z) = T_0 (1+z)
\label{eq:Tz}
\end{equation}
It is actually possible to measure the temperature of the background radiation
through the observations of the ratio of molecular lines: the ratio of the 
population on two levels for which the difference in energy is only a few
Kelvin is sensitive to the cosmic radiation field and  therefore provides a  way to actually measure the temperature of
the background. Such lines can be detected in the optical domain.
Actually the first detection of the CBR was obtained by this method with CN lines \cite{McKellar,NedCMB}.
It has also been successfully applied to distant QSO's to measure the variation o fthe temperature of the background with redshift and the result
are consistent with the expanding picture  \cite{Songaila1994,Srianand2008}. 

These measurements provide a fundamental test of the Big Bang picture but do 
not provide a source of constraint on the cosmological parameters.  \\

\begin{figure}
  \includegraphics[width=1.\textwidth]{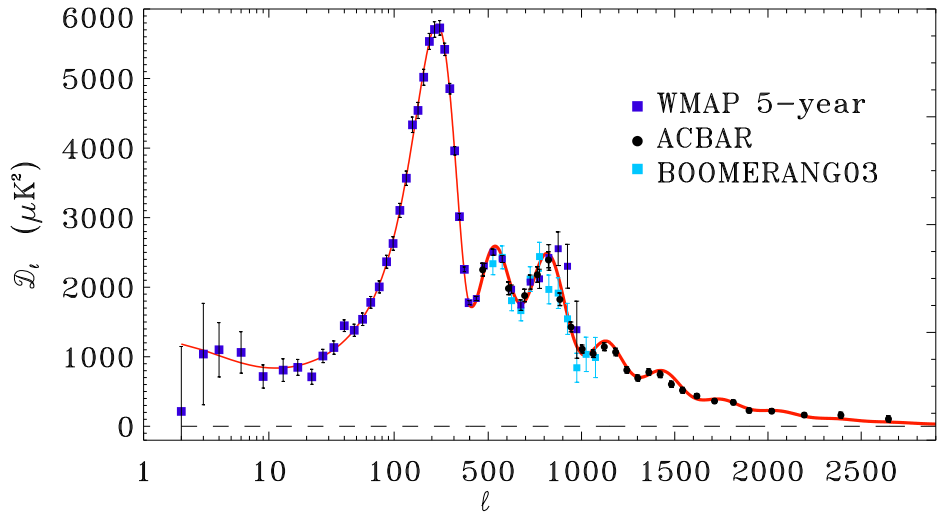}
\caption{ The amplitude of angular fluctuations of the CMB is expressed through their angular power spectrum. Data are  WMAP, Boomerang, ACBAR \cite{acbar2008}. A simple minimal six parameters model including a cosmological constant provides
an excellent fit to the data. This is one of the most important successes of modern cosmology. }
\label{fig:acbar}       
\end{figure}

\paragraph{ Fluctuations}
Since the discovery of the CMB fluctuations by COBE \cite{1992ApJ...396L...1S}
the idea that early universe physics has left imprints revealed by these 
fluctuations has gained an enormous attention. In this respect, 
DMR results have played a fundamental role in modern cosmology comparable
to the discovery of the expansion of the universe or the discovery  of the 
microwave 
background by Penzias and Wilson, and indeed this has motivated the delivering 
of the Nobel prize to G.~Smoot and J.~Mather for this discovery. One of the 
fundamental reasons for this is that  fluctuations
on scales larger than one degree in the microwave background radiation 
correspond to scales greater than the horizon at last scatering epoch 
and cannot therefore been altered by any physical process and should therefore reflect primordial fluctuations \cite{WeinbergI}. This also means that 
the very existence of these fluctuations 
could be explained only from yet undiscovered physics, probably relevant to the very early 
universe \cite{Guth}, for which the expansion law has been modified compared to the standard picture. The DMR results
were providing some constraints on cosmological models \cite{COBEpar} but 
it has been realized that the measure of fluctuations on smaller scales will
 provide  much stringent information. Early detection of fluctuations on 
degree scales allowed to set interesting constraints and provided the first 
evidence for a flat geometry of space \cite{Charley1,Charley2}. If 
estimations of low matter density were to be regarded as robust, this was inevitabily leading to a non-zero cosmologcial constant. Even before 
the availability of the WMAP data, considerable progresses have been achieved
on the measurement of fluctuations on all angular scales. Archeops \cite{Archeops} and Boomerang \cite{Boomerang}, as well as many other small scale measurements,
already provided measurements allowing tight constraints on cosmological parameters \cite{ArcheopsPar}. Although the observed fluctuations were consistent 
with a $\Lambda$ dominated universe, a cosmological constant was not 
explicitly requested by the CMB data alone. Indeed even the WMAP data 
were consistent with a vanishing cosmological constant, provided the Hubble 
constant was left as an entirely free parameter. A positive detection of a 
cosmological constant could be obtained only by using some additional  data in 
conjunction with CMB \ref{fig:acbar}, like  the measurement of the Hubble constant. 
A further restriction came from the fact that the constraint on cosmological 
parameters were obtained within the standard CDM picture, and that many 
ingredients were specified without being necessarily confirmed by 
 observations : for instance initial fluctuations are supposed to be adiabatic and to follow some power law. Therefore the ``concordance'' \cite{OstrikerSteinhardt} cosmology was an appropriate 
terminology: the model was consistent with most existing data, but the 
introduction of a cosmological constant was not requested by any single data, 
and it was far from being clear whether relaxing some of the input hypotheses
would not allow for solutions without the introduction of a cosmological constant.

\subsection{Clusters of galaxies}

Most talks on clusters of galaxies start with  stating the fact that they are 
the largest virialized structures in the universe. Another reason for the strong
 interest they represent is due to the fact that clusters are the only 
astronomical large objects for which we have  such a wide range of information:
 their total mass is in principle accessible from x-ray spectroscopy, velocity 
dispersion, weak lensing signal, their gas content can be investigated through 
x-ray observations but also through their imprint on the microwave 
background, the thermal Sunyaev-Zel'dovich effect.  Obviously their star content
 can be evaluated, but also the metal content of their gas. Clusters are 
therefore the astrophysical objects relevant to cosmology on which the most comprehensive set of 
data can be obtained.

\subsubsection{Baryon fraction}

X-ray data allow to determine at the same time the mass of the gas content of 
clusters, through imagery,  as well as the gas temperature., through spectroscopy. The temperature can be used  to estimate their total mass. It is therefore straightforward then to infer 
their 
baryon fraction (most of the baryons being in the gas, the star even if uncertain can be taken into account without making much difference). Because, the 
gravitational collapse is expected to be identical for dark matter and for 
baryons, at least before the first shocks occur, the baryon fraction should 
reflect the primordial baryon fraction~:
\begin{equation}
\frac{M_b}{M_t} = \Upsilon  \frac{\Omega_b}{\Omega_M}
\label{eq:fb}
\end{equation}
where $\Upsilon$ is expected to be close to one. Numerical simulations shows 
indeed that $\Upsilon$ should lie in the range $0.85 - 1.$. 
${\Omega_b}$ can be known from primordial nucleosynthesis results, and therefore constraints can be obtained on $\Omega_M$ \cite{White1993}. 
Although this test is attractive in its principle, its application, especially 
in the context of precision cosmology, is somewhat problematic. There are 
several reasons for that: for instance,  in numerical simulations, as well as in the observations, 
$\Upsilon$ seems to depend on the radius and is therefore a function $\Upsilon(r)$  
on radius. The precise value of $\Upsilon(r)$ is likely to depend on the physics of the 
 non-gravitational heating and cooling processes which are necessary to reproduce observed properties of clusters in numerical simulations. However, 
these possible limitations still allow to infer an upper limit on $\Omega_M$ from eq. \ref{eq:fb}. A final limitation comes from the clumping of the 
gas which would lead to an underestimation of $\Omega_M$.
On the observational side, it is unclear to 
which accuracy cluster masses are  
known~: agreement between lensing masses and x-ray mass is under debate. 
Estimations of total mass and baryon mass in the outer parts of clusters, 
which are expected to provide the most reliable estimation of the primordial 
baryon fraction, are subject to uncertainties: the x-ray emission 
of the gas at the virial radius is low and only for few clusters 
do we have a firm detection. At this outer part, the gas is eventually clumped
which would bias the gas mass estimates. 

\subsubsection{Geometrical tests}

Another strategy is to use clusters as standard candles in some sense 
and to constrain the cosmological parameters from the observed properties of distant
 clusters. For instance it has been proposed to use cluster x-ray radius  as a 
standard ruler  \cite{Mohr}. Additional information may allow direct distance measurements.
This is what the Sunyaev-Zel'dovich (SZ)  signal provides: because the x-ray 
emissivity goes as the square of the gas density, while the SZ signal just goes
 as the density the distance can be directly obtained. Not only this method 
can be used to estimate the Hubble constant, it is also possible to 
extend the Hubble diagram to distant clusters and thereby to get the same 
information provided by the SNIa diagram.  Another variant of these tests has 
been proposed from the baryon fraction of clusters 
against redshift: because the gas fraction inferred from observations depends 
on the angular distance to the cluster, requesting that the gas fraction does 
not evolve with redshift will result on constraints on the cosmological parameters
\cite{Sasaki}.
This test has deserved recent attention \cite{Allen2008}, but uncertainties mentioned about the 
baryon fraction are likely to alter this test either and it is difficult to 
draw robust conclusions from it \cite{sadat05,luisfb,Ettori2009}.

\subsubsection{The Use of Cluster Abundance}

\paragraph{The theoretical mass function}

A theoretical expression for the abundance of virialized structures in 
hierarchical picture  has been first proposed by Press and Schechter in 1974 
\cite{PS}. From scaling arguments, a general expression for the mass function can be derived \cite{bvm}:
\begin{equation}
 N(M,z)  = - \frac{\overline{\rho}}{M^2\sigma_z(M)} \delta_{NL}(z) \frac{d\ln \sigma \;}{d\ln M}{\cal F}(\nu) 
\label{N(m)}
\end{equation}
$\delta_{NL}(z)$ being the threshold for non-linear collapse of  spherical density
 perturbations (in an EdS universe $\delta_{NL}(z)= 1.68$),  $\sigma_z(M)$ the amplitude of mass fluctuations on the mass 
scale $M$ ($\sigma_z(M)= \sigma_0(M)D(z)$, $D(z)$ being the normalized growing rate of linear fluctuations) and $\nu_{NL}$ the normalized threshold ($\nu_{NL}=\delta_{NL}/\sigma(M)$
) for non-linear collapse, $\int_\nu^{+\infty}{\cal F}(\nu)d\nu$ is the fraction of space covered by spheres of mass $\geq M$ satisfying the non-linear criteria. Such an expression can be rigorously justified only for a given power law 
spectrum in an Einstein-de Sitter universe. However the numerical investigations
of the mass function have shown that it does follow the above scaling to a very high accuracy \cite{MassFunctionEfstathiou,ST,Jenkins}.  Press and Schechter used for the function $\cal F$:
\begin{equation}
 {\cal F}(\nu) = \sqrt{\frac{2}{\pi}}\exp (-\frac{\nu^2}{2})
\end{equation}
while the expression proposed by Steh and Tormen \cite{ST,ST2}, possibly extending the non-linear condition to take into account ellipsoidal collapse,  was found to produce a more  accurate agreement with numerical simulations~: 
\begin{equation}
{\cal F}(\nu)  = \sqrt{\frac{2 A}{\pi}}C\exp (-0.5A\nu^2)(1.+ (1./(A\nu)^2)^Q)
\end{equation}
with  $A = 0.707$ $C = 0.3222$ and $Q = 0.3$. Jenkins et al. \cite{Jenkins} did provide another fitting formula in which the amplitude of $\delta_{NL}$ is set constant (independently 
of the cosmology) which was found to provide an accurate fit to within 20\% in 
relative amplitude\footnote{Another complication happens from the definition of
 ``an object with a mass $M$''. Two algorithms are commonly used, the friend of
 friend one (FOF) and the spherical overdensity one (SO). In addition the non 
linear contrast density $\Delta$ which is used could be set to an arbitrary 
fixed value or to depend on Cosmology according to the virial value 
$\Delta_v$ evaluated within the spherical collapse model.}. This accuracy is sufficient for 
present-day available data to be compared with. More recently larger numerical simulations were used \cite{Warren,Tinker2008,Crocce} to investigate the mass function to a precision higher than 20\%, finding some departures from strict universality and provide more accurate fitting formula,  but cosmological data are not yet at a precision level for which such departures could be observed.

\paragraph{Local abundance}  

The observed local abundance of clusters of a given mass $M$ can be used to 
infer directly the average amplitude of matter fluctuations, $\sigma(M)$, through the 
expression of the mass function. Such a derived  value  depends  weakly 
on Cosmology \cite{BB} as Cosmology enters only through  $\delta_{NL}$, which as we have 
mentioned is almost independent on  Cosmology. However, the traditional way 
to express the amplitude of matter fluctuations is through $\sigma_8$ and 
therefore the dependence of this quantity  appears on Cosmology and on the 
initial spectrum. There is a source  of uncertainty  
 due to the fact that the mass is not directly an observable 
quantity. Therefore one has to use a relation between  mass and some 
observable quantity. At the same time, the selection function of clusters
 within a given sample has to be known in order to infer the actual number 
per unit volume.  X-ray clusters are a priori the easiest choice, as the 
selection function from x-ray surveys is easy to model and the x-ray 
temperature is a priori a 
reliable and physically motivated proxy of the total mass. There are various arguments from which one expects that $\displaystyle T_x \propto {GM}/{r}$. This leads to a scaling relation between mass and temperature:
\begin{equation}
 T =A_{TM} M^{2/3}_{15}(\Omega_M (1+\Delta)/179)^{1/3}h^{2/3}(1+z)\rm \;\; keV
\label{eq:mt}
\end{equation}
an expression in which the contrast density $\Delta$ is the relative density compared 
to the background density of the universe. An other density contrast is also 
used which refers to the contrast  density $\Delta_c$  relative to the critical density of the universe $\rho_c(z)$. The above scaling reads then:
\begin{equation}
 T =A_{TM} M^{2/3}_{15}(\Omega_M (1+\Delta_c)/179)^{1/3}(E(z)h)^{2/3}\rm \;\; keV
\end{equation}
where the redshift dependent Hubble constant is written $H(z) = 100 h E(z) \rm km/s/Mpc$. There has been much debate on which value of the normalization 
$A_{TM}$ is to be used and this translates in an uncertainty in $\sigma_8$ 
\cite{Pierpaoli}. The various values published for $\sigma_8$, ranging from 
0.6 to 1.05 for a $\Lambda$CDM model with $\Omega = 0.3$ \cite{ATM} are 
consistent with each other when the various values of the normalization constant $A_{TM}$ are taken into account. However, because of this uncertainty 
the actual value  of $\sigma_8$ inferred from cluster abundance is still
 subject to some debate \cite{Reiprich2006,Evrard2008,Rines2007}.

\paragraph{Abundance evolution}

Peebles et al. \cite{Peebles89} and Evrard \cite{Evrard1989} discussed the constraints that the existence of high 
redshift clusters were setting on Cold Dark Matter models, at a time when 
Einstein-de Sitter Cosmology was fashionable. These works followed some earlier ones on the use of cluster abundance in cosmology \cite{Perrenod}. 
The 
evolution of the abundance of clusters  of a given mass is a sensitive function 
of the growing rate of fluctuations  therefore offering 
 a powerful 
cosmological test   \cite{Lilje,OB92}. Rather obviously, this test needs that local abundance to   be adequately normalized.  
This test is primarily sensitive to the cosmological density of the Universe, very weakly depending on other 
quantities   \cite{BB}.  
Attempts to apply directly the test of the evolution of the temperature distribution function  of
clusters have been performed but still from a 
very limited number of clusters (typically 10 at redshift 0.35) and lead to somewhat apparently conflicting results \cite{H97,VL,B2000}.\\

\begin{figure}
\includegraphics[width=\textwidth]{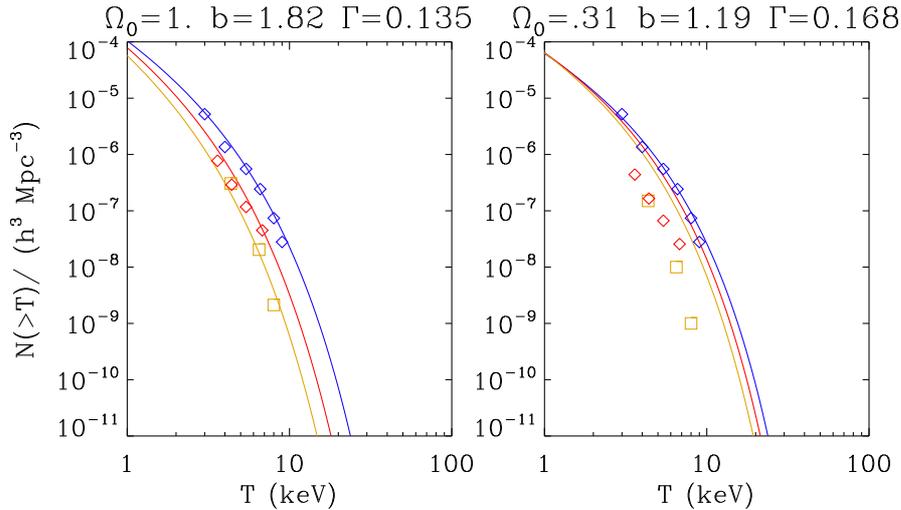}
\caption{ The low and high redshift temperature distributions compared to predictions of  an Einstein de Sitter Universe (left side, $\sigma_8 = 0.55$, $A_{TM} = 6.35$ in eq. \label{eq:mt}, $\Gamma = 0.135$ is a shape parameter for the spectrum) and to predictions of the concordance model (right side, $\sigma_8 = 0.81$, $A_{TM} = 7.75$, $\Gamma = 0.168$ ). Upper (blue) losanges are the local data. Lower (red) losanges are the temperature distribution function ($z \sim 0.33)$) derived from EMSS, (yellow) squares  are high redshift ($z \sim 0.5)$) clusters from MACS and the 400 sq deg survey. }
\label{fig:n(T)}       
\end{figure}

  On the 
other hand,
 redshit number counts (without temperature information available) 
allow one to use samples comprising many more clusters.
 Indeed using simultaneously different existing surveys one can use information provided by more than 300 clusters with $z > 0.3$
(not necessarily independent). 
 In order to model clusters number counts, for 
which temperatures are not known, it is necessary to have a good knowledge of
the $L-T$ relation over the redshift range which is investigated \cite{OB97}.
Early investigations indicated the absence of evolution of the $L-T$ relation or a slight positive evolution, indicative of a high density universe \cite{S97,R99,Borg99}. The $L-T$ relation at high redshift has been more recently 
determined  by the XMM-$\Omega$ project \cite{OP,OP1} and Chandra \cite{V02}. Number counts can then be computed:
\begin{eqnarray}
N(>f_x,z,2\Delta z) = & \Omega \int_{z-\Delta z}^{z+\Delta z} \frac{\partial N}{\partial z}(L_x> 4\pi D_l^2 f_x) dz \nonumber \\
= & \Omega \int_{z-\Delta z}^{z+\Delta z} N(>T(z))dV(z) \nonumber \\
= & \Omega \int_{z-\Delta z}^{z+\Delta z} \int_{M(z)}^{+\infty}
N(M,z)dM dV(z)
\end{eqnarray}
 where $T(z)$ is the temperature threshold at redshift $z$ corresponding to the flux $f_x$
as given by the observations, being therefore
independent of the cosmological model.
This approach has been used to model all available ROSAT surveys and using the latter  
$L-T$ measurement provided by XMM \cite{Vauclair03}.
 All  existing x-ray cluster surveys systematically point toward high 
$\Omega_M$. A combination  allowed a determination of   $\Omega_M$ with a 15\% precision: $ 0.85 < \Omega_M < 1.15 {(1 \sigma)} $ (depending somewhat on the calibration issue, this is part of the systematic). 
During this analysis numerous
 possible sources of systematics were investigated with great detail  and are 
 representing roughly an additional  15\% uncertainty. This means that 
global uncertainty is roughly 20\%. This gives unambiguous evidence that the 
observed high redshift cluster abundance compared to local one is inconsistent with the one modelled within the concordance model. 
This is confirmed by the recently availability of a large sample of high 
redshift clusters with measurements provided by Chandra \cite{Vik09} and a 
sample of high redshift massive clusters  \cite{Ebeling07}. The temperature 
distributions drawn from these samples is shown in figure \ref{fig:n(T)} : in agreement  with previous 
results
these temperature distributions are inconsistent with the concordance cosmology.
A critical ingredient in the modeling of x-ray clusters  is the  mass temperature relation \ref{eq:mt}. A possible strong evolution of this relation, as proposed to reconcile redshift number counts with predictions \cite{Vauclair03}, would allow the agreement  of the concordance model with the observed temperature distributions as well. 

\subsection{Large scale Structure of the Universe}

\begin{figure}
\includegraphics[width=0.75\textwidth]{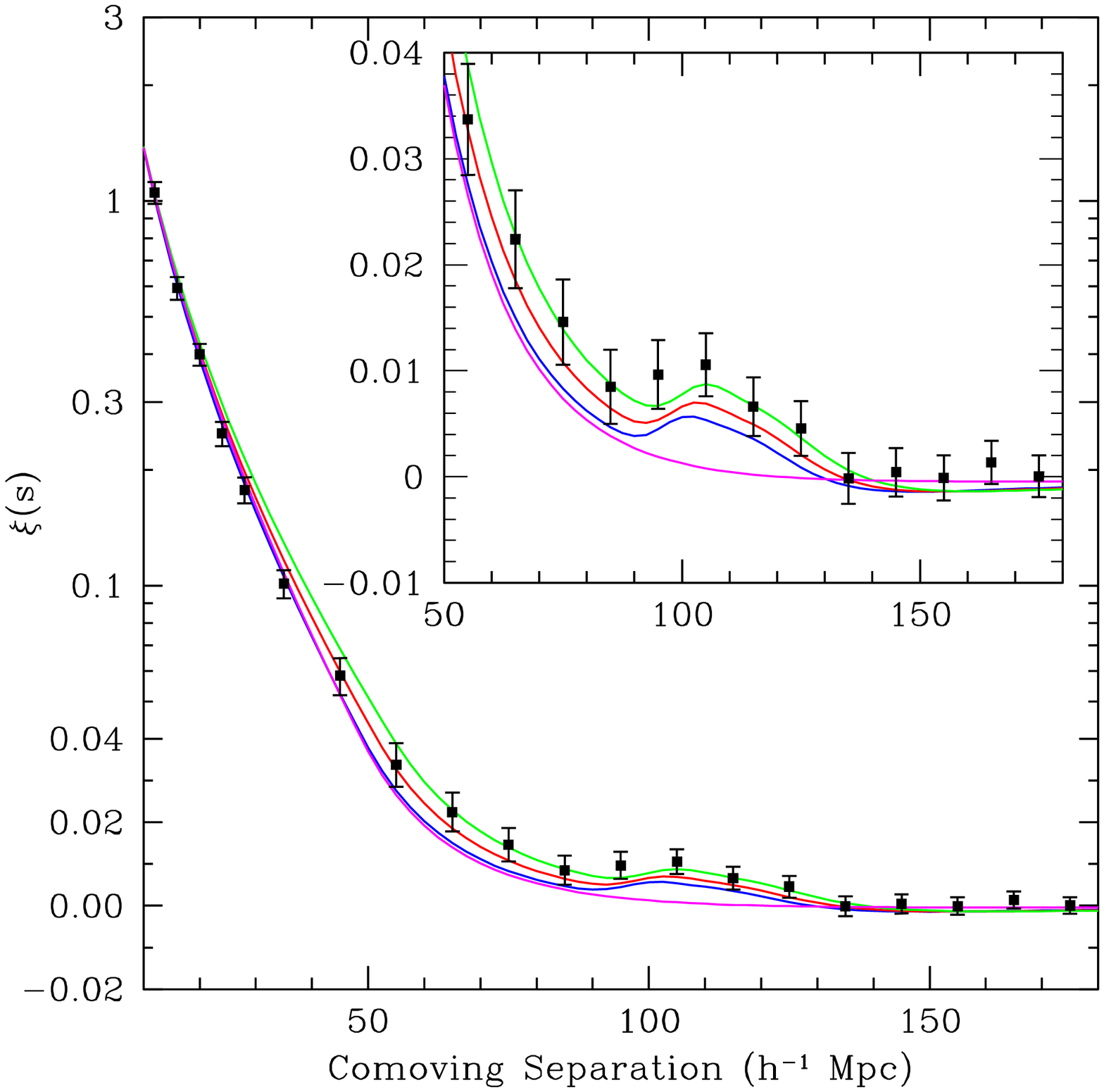}
\caption{ Large scale correlation function of the SDSS Luminous Red Galaxies \cite{Eisenstein2005}.
Three $\Lambda$CDM models ($\Omega_Mh^2 = 0.12,0.13,0.14$) are shown with a baryonic content evaluated from the 
CMB ($\Omega_bh^2 = 0.024$). A model without baryon corresponds to the bottom (magenta) line.   }
\label{fig:xiSDSS}       
\end{figure}

The properties of the galaxy distribution on large scales has been a major 
source of constraints for cosmology in the past. The most often quoted quantity
is the correlation function of galaxies. Until the 1990's the correlation function was relatively well know only on scales which are non-linear or nearly so.
Therefore the correlation function could be adjusted only by  means of numerical simulations. Two major breakthroughs enabled important progresses in this area~:
the first one was the measurement of the correlation function of galaxy from the APM survey 
allowing to exclude the standard cold dark matter picture \cite{MaddoxAPM}.
Furthermore the data were consistent with the predictions of a low density 
CDM model with $\Omega_M \sim 0.3$ and it has been argued \cite{Efstathiou90}
that the introduction of a cosmological constant was a way to reconcile CDM with inflation predictions, in the spirit of an earlier proposition \cite{Peebles84}. The second one was the discovery
 of a simple analytical formalism allowing to construct the correlation 
function
in  the non-linear regime from the initial spectrum \cite{hklm}. More recently, the most 
critical advance resulted from the availability of very large galaxy surveys, the 2Df redshift survey and the SDSS survey, allowing 
to measure the amplitude of galaxy fluctuations on scales as large as $100h^{-1}$ Mpc \cite{Percival01,Tegmarketal2004,Eisenstein2005,Percivaletal2007}. This has provided a remarkable success to the $\Lambda$CDM picture because the amplitude of the correlation function could be predicted for models that 
already match the CMB fluctuations measured by WMAP: not only $\Lambda$CDM model reproduces the shape of the correlation function, but the specific presence 
of a bump in the correlation function at scale of the order of $100h^{-1}$Mpc due to the detailled dynamics of fluctuations when the baryons are taken into account, the so called accoustic peak, corresponding to the ``peak'' in the ${\cal C}_l$ of the CMB (Fig. \ref{fig:xiSDSS}). More detailed use of pwoer spectrum measurements might be limited by our understanding of the exact relation between galaxies distribution and 
the underlying dark matter distribution \cite{SanchezCole}.


\section{Possible origin of  the apparent acceleration}
\subsection{An Einstein-de Sitter Model ?}

\paragraph{Supernovae Hubble diagram}

\begin{figure}
  \includegraphics[width=0.75\textwidth]{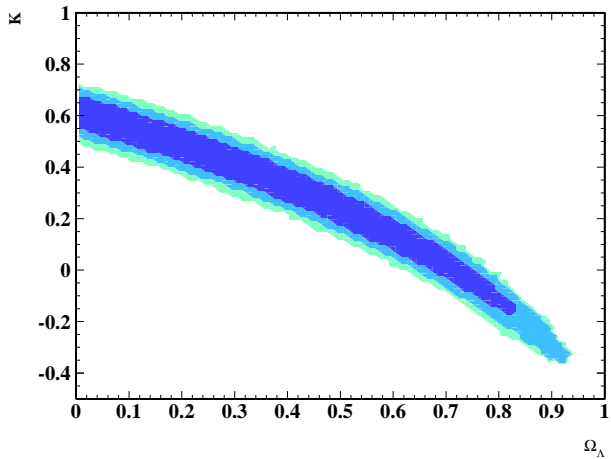}
\caption{Fitting the SNIa Hubble diagram with two free parameters, one being the cosmological constant  in a flat cosmological model and the second being a parameter describing a possible time evolution of the luminosity of distant supernovae ($\Delta m(z) = K (t_0-t(z))$) leads to the following constraints \cite{lfabyz}. 
Contours are 1, 2 and 3 sigma regions. There is a severe degeneracy between the two parameters which prevents an unambiguous determination of the cosmological constant. A vanishing cosmological constant is entirely consistent with SNIa diagram provided that a significant but not unrealistic amount of evolution is allowed for. The major issue is that it is
extremely difficult from a purely observational point of view to reject such a possibility. }
\label{fig:1}       
\end{figure}

The first convincing evidence for acceleration is generally considered as coming
from the SNIa Hubble diagram. However, use of geometrical tests based on the assumption 
of no-evolution of the parent population of the test is always 
possibly subject to produce biased values because of un-anticipated evolution.
One possible way to cure this problem is to assume some evolution and see 
whether the data still provide evidence for the claim. For instance, an evolution term like:
\begin{equation}
\Delta m_e \propto z
\label{}
\end{equation}
can not mimic the observed Hubble diagram. 
The limitation of this 
approach is due to the fact that we have no model for this evolution 
and therefore we are left with a purely empirical approach. Other forms of evolution may therefore lead to different conclusions. Indeed, it has been suggested that the correcton term might be:
\begin{equation}
\Delta m_e \propto \Delta t
\label{eq:evsn}
\end{equation}
\cite{wrightSN}, such term leads to  large degeneracy between cosmology and possible evolution \cite{lfabyz,Linden} (Fig. \ref{fig:1}).\\
Undoubtfully, despite its possible limitation, the determination of the Hubble diagram from  SNIa has led to a 
major and rapid change of paradigm in modern cosmology. However, this change 
has been possible because the previous situation was problematic. Although 
some observational  indications were favoring a low density universe, the first 
detections of fluctuations on degree scales were in conflict with open low density universe \cite{Charley2}.

\paragraph{ Fluctuations of the Cosmological background radiation}

\begin{figure}
\centering
\includegraphics[height=8cm]{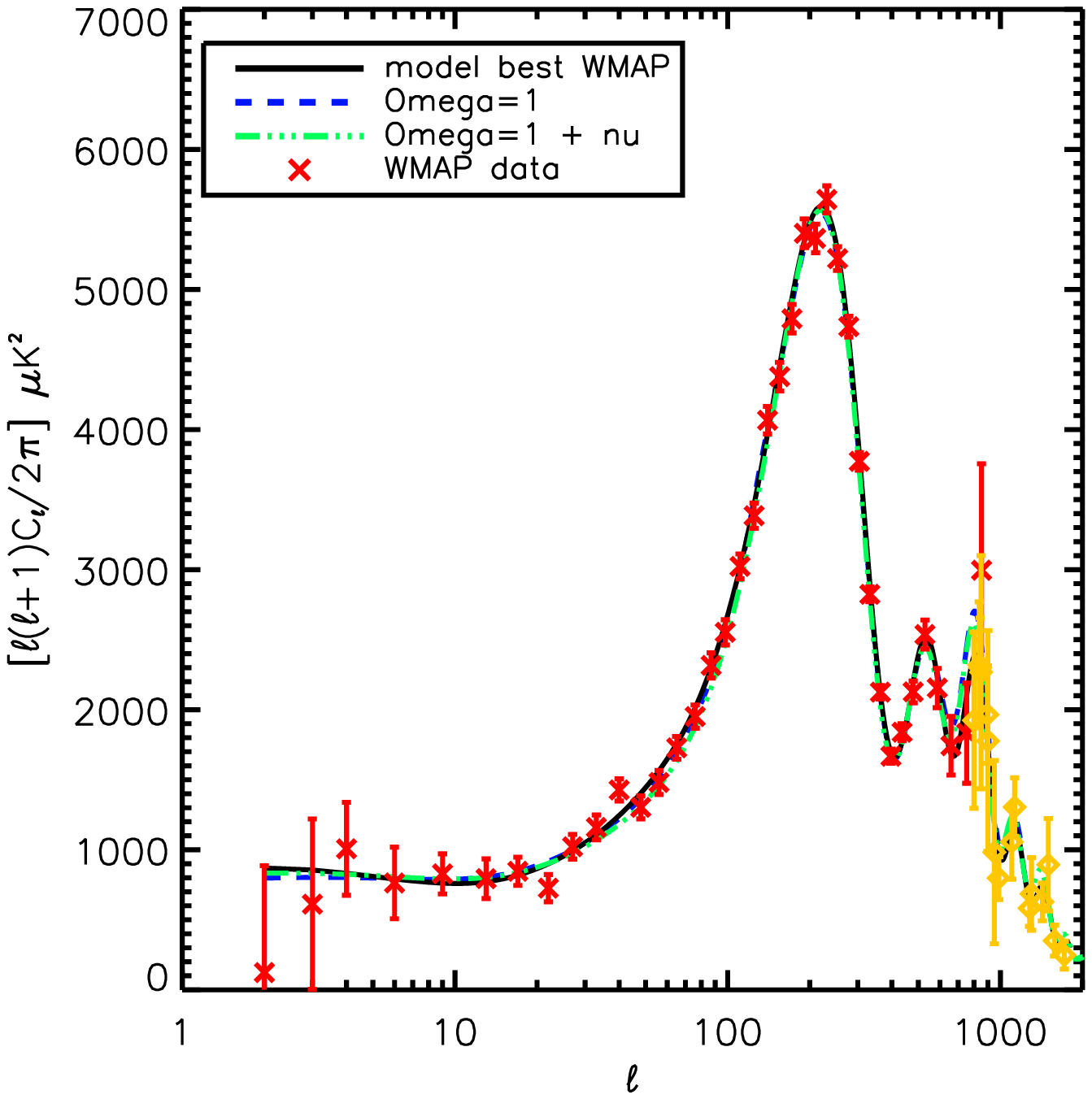}
%
%
\caption{The TT spectrum of the first year WMAP data compared to three different models: 
one is the concordance, the two others are Einstein de Sitter models, one of 
which  comprises neutrino contribution of $\sim 10\%$ corresponding to three degenerate families with $m_\nu \sim 0.7 $eV. From Blanchard {\it et al.} \cite{2003A&A...412...35B}.}\label{fig:dg}
\end{figure}
 
The remarkable results of the WMAP 
experiment, with accurate measurements of the ${\cal C}_l$ and additional measurements on the polarization, are often quoted 
as providing a direct evidence for an accelerating universe. This is 
incorrect: cosmological constraints  established from CMB entirely rely on the spectrum shape assumptions, which is commonly assumed to be described by a single power law. 
Therefore the conclusions on the high precision obtained on cosmological parameters could be erroneous
\cite{Kinney2001}. 
Indeed, relaxing this hypothesis, i.e. assuming a non power law power spectrum, it is possible   to 
produce ${\cal}C_l$ curve within an Einstein de Sitter cosmological model 
which provides a fit as good as the concordance model. This is illustrated
in figure \ref{fig:dg} on which 3 models are compared to the WMAP data, two 
being 
Einstein de Sitter models. Such models not only reproduce the TT (temperature-temperature) spectrum,
but are also extremely close in terms of ET (polarization-temperature) and EE (polarization-polarization) spectra. Furthermore 
the matter power spectra are similar on scales probed by current galaxy 
surveys before the availability of the SDSS LRG sample. An un-clustered component of matter like a neutrino contribution or
a quintessence field with $w \sim 0$ is necessary to obtain an acceptable
amplitude of matter fluctuations on clusters scales \cite{2003A&A...412...35B}.
Such models require a low Hubble constant $\sim 46$ km/s/Mpc. Such a value
might be viewed as terribly at odd with  canonical HST key program value ( $\sim 72$ km/s/Mpc) but is actually only $\sim  3 \sigma$ away 
from this value, this can 
certainly not be considered as a fatal problem for an Einstein-de Sitter 
universe. The introduction of a non-power law power spectrum might appear as 
unnatural. However, such a feature can be produced by some models
of inflation in order to match the ${\cal C}_l$ curve \cite{HuntSarkar}. Therefore the amplitude and shape of the 
CMB fluctuations as measured by WMAP is certainly a success for  the $\Lambda$ CDM model but cannot be regarded as a direct indication of the presence of dark energy.

\paragraph{Large scale structure}

\begin{figure}
\centering
\includegraphics[height=7cm]{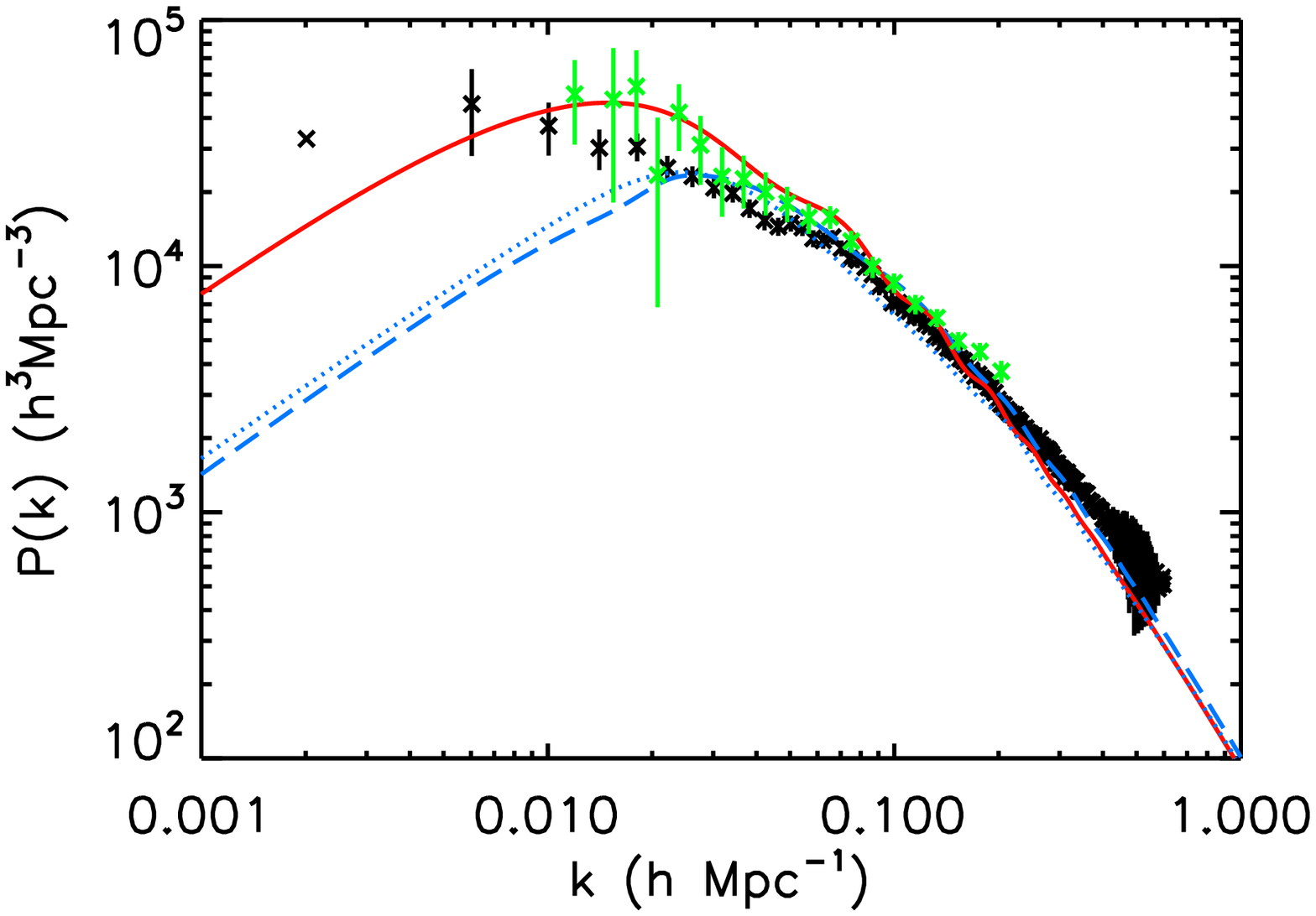}\vspace*{10mm}
%
%

\caption{Data from the SDSS have allowed to measure the amplitude of galaxy fluctuations on large scales. In this respect, Luminous Red Galaxies (LRG) provided 
measurement of the power spectrum on the largest scales. Green crosses correspond to Tegmark et al. \cite{Tegmarketal2006} and black crosses correspond to the latest 
measurements of the power spectrum of LRG from the SDSS Data Release 5 by Percival et al. \cite{Percivaletal2007}.   The red continuous curve is the predicted spectrum for a typical concordance model, while the dotted and dashed lines correspond to the power spectrum for Einstein de Sitter models consistent with the
WMAP fluctuation angular power spectrum ${\cal C}_l$ \cite{2003A&A...412...35B,HuntSarkar}.  }
\label{fig:pk_lls}
\end{figure}

Once an Einstein de Sitter model is built in order to reproduce the CMB ${\cal C}_l$,
the amplitude of the matter fluctuations  on large scales is set up and the 
measurement of the matter fluctuations on large scales in the present day universe 
is a critical way to distinguish models which are otherwise degenerated in
their ${\cal C}_l$. The comparison of the power spectrum from the SDSS LRG with the
 predicted spectra for Einstein de Sitter models is clearly in favor of the concordance
 model \cite{2006A&A...449..925B}, see Fig. \ref{fig:pk_lls}. One should add some caution here: it might be possible that the biasing 
mechanism leads to a power spectrum at small $k$ (large scales) which is not 
proportional to the actual matter power spectrum \cite{Durreretal2003}, in 
which case the above comparison might not be a fatal failure of the Einstein 
de Sitter models. However, biasing mechanisms systematically lead to a 
correlation function on large scales which is still proportional to the 
matter correlation function on large scales. Comparison 
of the correlation function on large scales is therefore less ambiguous
and its measurement  should be unambiguously discriminant. 
Hunt and Sarkar \cite{HuntSarkar} have provided a comprehensive MCMC investigation of 
the Einstein de Sitter parameter space, finding models which acceptably 
fit the correlation function on scales below 70 $\rm h^{-1}Mpc$, but were 
nevertheless systematically negative on scales of the BAO peak. This is a 
strong evidence that there is no way in an Einstein de Sitter universe 
to fit  simultaneously the ${\cal C}_l$ and the observed distribution of galaxies on large scales. This should be regarded as a remarkable success of the 
concordance cosmological model: although there were little doubts that
this model could fit accurately most of the major existing 
observational facts in  cosmology, the ability to produce predictions 
that are verified a posteriori is the signature of a satisfying scientific theory. Number counts of x-ray clusters were found to match  the expectations 
of  an EdS Universe and conflict with the concordance model. However, if the standard scaling of the $M-T$ relation 
\ref{eq:mt} is broken by some non-gravitational processes, the number counts and 
temperature distribution function could be altered in a way that the concordance model can accomodate data \cite{Vauclair03}.

\subsection{Einstein Cosmological constant or Vacuum  contribution}

The most direct explanation one can provide to the cosmic acceleration it that it 
is 
 due to a true cosmological constant appearing in the geometrical part
of Einstein's equation, i.e. the left hand side of eq. \ref{eq:Einstein}. However, it is much more common to believe that 
this term arises from some contribution to the energy-momentum tensor on the 
right hand side. 
As we have seen, from a classical point of view the vacuum 
might have a non zero density and behaves identically to a cosmological 
constant. In addition, quantum mechanics  provides an intriguing hint in this 
direction. The possible energy levels of an harmonic oscillator  are known to be:
\begin{equation}
E_n = (n+\frac{1}{2})h\nu
\label{eq:oh}
\end{equation}
and so the state of lowest possible energy, the zero point,  is not zero 
but $h\nu/2$.  This non-zero value is often noticed in standard text book of quantum mechanic, but,
because observable quantities 
correspond to transitions from one state to an other one, is not regarded as being problematic. However, as soon as 
gravitational interaction has to be added, one cannot avoid to take the absolute energy into account. Summing all the contributions of modes of the fluctuations of the electromagnetic field  up to some wave number $k_c$ gives a density $\rho_V$:
\begin{equation}
  \rho_V = \int_0^{k_c} \frac{4\pi k^2dk}{8\pi^3}\frac{1}{2}k\sim \frac{{k_c}^4}{8\pi^2}
\label{eq:rhovkc}
\end{equation}
The total contribution therefore depends on the cutoff scale $k_c$. If the Planck scale is taken, this leads to a density which is something like $10^{120}$ too large. For all  energy scales in physics does $\rho_V$ end up with an unacceptable large value, which looks like a fundamental problem \footnote{It has been suggested that the zero-point fields should not be 
regarded as real, despite the fact that they are at the basis of the calculation of 
the Casimir effect \cite{Michel}.}.   An elegant solution to this problem is obtained from supersymmetry: the contribution to vacuum from fermion is negative and therefore with an equal number of modes in fermions and bosons one gets a cancellation. However, because the supersymetry should be broken at energy below 1 TeV or so,
the problem of the vacuum density is still not solved, even if its strength has been noticeabily reduced.  An other proposition is that the vacuum actually behaves differently from \ref{eq:rhovkc}\cite{Branchi2}.

 \paragraph{Detecting  the cosmological constant in a laboratory experiment?}

The above considerations might lead to a fascinating possibility. 
If the observed dark energy corresponds to an actual contribution of the 
electromagnetic field there will be a frequency $\nu_c$ associated to the 
cut off in \ref{eq:rhovkc}, and the value associated with the  present value of dark energy is:
\begin{equation}
\nu_c \sim 1.7 \ 10^{12} \rm Hz
\label{eq:nucut}
\end{equation}
Vacuum fluctuations produce noise which can be detected experimentally 
with Josephson junctions \cite{Koch}, it is therefore possible that the 
frequency cut-off $k_c$ could be measured in a laboratory experiment 
\cite{Beck&M}. This point of view has been criticized by 
\cite{Jetzer&Straumann}. A further problem arises from the fact that the 
existence of a cut-off like \ref{eq:nucut} could modify the classical Casimir 
force\cite{Casimir}, in a way which is excluded by existing data \cite{Mahajan-et-al2006}.
However, the origin of the noise current in the Josephson junction may have nothing to do with the electromagnetic
field in the device \cite{Branchi1}.  Others arguments to reject this option have proposed (\cite{CasimirLambda}).

\subsection{The effect of inhomogeneities}
\label{inhom}
 The standard formalism of Friedmann-Lema\^\i tre models relies on 
the assumption that matter is uniformly distributed. Since the early work
of Sachs and Wolfe  \cite{SachsWolfe},  perturbations are commonly written as departure 
$h_{\alpha\beta}$ from the Robertson-Walker metric element $g_{\alpha\beta}$:
\begin{equation}
\tilde{g}_{\alpha\beta} = g_{\alpha\beta} + h_{\alpha\beta}.
\label{eq:perturb}
\end{equation}
This is a qualitative expression, as there are many ways to write perturbations
depending first on the choice of the initial form of the metric and  coordinate systems. 
The analysis of linear perturbations  ($\vert h_{\alpha\beta}\vert \ll \langle g_{\alpha\beta} \rangle$) is rich and not as trivial as one might naively think. However,  
the dynamics of relativistic linear perturbations to first order 
is well understood. Three distinct physical modes are possible:
scalar perturbations
which represent fluctuations of the density field $\rho$, vector perturbations
which represent  the vorticity of the velocity field  and tensor perturbations which 
represent gravitational waves.  Only this last term is obviously not present 
in Newtonian dynamics.  This has legitimated the use of Newtonian theory to describe evolution and average effect of perturbations on 
scales much smaller than the Hubble scale and for velocities much smaller 
than the speed of light. It is therefore common to use the  
Newtonian perturbed RW metric:
\begin{equation}
 ds^2 = -(1+2\psi) dt^2 + a(t)^2\left(1-2\psi\right)\gamma_{ij}dx^idx^j
\label{eq:conformNew}
\end{equation}
$\psi$ being the Newtonian potential. However, the fact that we live in an inhomogeneous 
universe has always left open the possibility that average observable 
quantities may be definitively different from what can be obtained in a 
rigorous homogeneous world, this question discussed in the context of relativistic world models, can be traced back to thirties \cite{Eddington1930,Tolman1934}.
One early worry of this kind was about the 
average effect of gravitational lensing. It has been suggested that the 
average magnification from inhomogeneities in the Universe might produce 
an apparent redshift-magnitude relation which behaves differently from 
that in a  homogeneous universe and 
would therefore bias inferred values of the cosmological parameters \cite{DyerRoeder}.
S.Weinberg \cite{Weinberg76} did provide a general argument to show that 
there is an integral constraint that guarantees the average relation in
inhomogeneous worlds to be  identical to the homogeneous case but that 
the way perturbations are handled may lead to lose this general integral 
constraint on average quantities even to the first order in perturbation 
(while it is valid at all orders).  Although Weinberg's theorem is extremely 
clever, it relies on some hypotheses, one is that the universe still behaves as a FLRW model with its density being equal to  the average density, and the subject has still been addressed
in recent years \cite{HolzWald,KibbleLieu2005,Barausse2005,BiswasNotari2008}.

In this approach, the consequences of the presence of inhomogeneities were analysed assuming the background evolution of the Universe was still following
 the usual Friedmann-Lema\^\i tre equations. However, beyond the problem of properly evaluating the observable consequences of the presence of inhomogeneities (as illustrated by the calculation of the fluctuations of the CMB), one fundamental 
question is whether the Friedmann-Lema\^\i tre equations describing the 
expansion could be significantly altered by the presence of inhomogeneities,
an effect named ``back reaction'' \cite{Futamase1989}.
At first look one would naively expect that the average effect of linear perturbations to first order is zero and that to the second order one would get something like (for the case of perturbations around an Einstein de Sitter universe):
\begin{equation}
\left(\frac{\dot a }{a}\right)^2 = \frac{8\pi G\rho }{3}   \left(1 + F(h_{\alpha\beta})\right)
\label{}
\end{equation}
with :
\begin{equation}
 F(h_{\alpha\beta}) \propto \ \langle h^2\rangle
\label{}
\end{equation}
For the  observable part of the universe, as we are within some  perturbation with a typical size corresponding to the horizon,  one expects that 
\begin{equation}
\rho_{obs} \sim \rho \left(1 \pm h_{\alpha\beta}\right)
\label{}
\end{equation}
(this is written in a very loose way that would probably horrify relativity 
experts, but it is done for the purpose of a qualitative illustration on how the 
problem is settled). Now astrophysical data show that $\langle h\rangle$ is always 
of the order of $10^{-5}$, or less, except in the vicinity of neutron stars and black 
holes, departures at this level would therefore not be very surprising. Actually the non linear collapse of structures like clusters or galaxies
 modestly enhances initial metric perturbations by a factor $\Delta^{1/3}$, but lensing 
measurements of galaxies and clusters directly prove that on these scales the 
metric perturbations  remain tiny and one expects the linear approximation to
 be sufficient even when the structures are becoming non-linear ($\frac{\delta \rho}{\rho} \gg 1$). Of course it is 
clear that some observables would be distorted at a level of $h \sim 10^{-5}$. However, the possible non negligible 
contribution of inhomogeneities to the global dynamics of the present universe, the so called ``back reaction'' has been 
proposed\footnote{The idea that the small scale evolution may have a connection
with  large scale dynamics in relativistic cosmology can be traced back to Eddington in 1930 \cite{Eddington1930}} with the idea that the actual dynamics of 
the universe is much more distorted by non-linear terms than naively anticipated. Ironically enough, early propositions of back reaction were suggested 
as a way to reduce or even cancel the presence of a large cosmological 
constant \cite{earlyBR,Mukhanovetal1997,Brandenberger2002}, while nowadays, the important 
question is whether this back reaction could explain the apparent cosmological 
constant. This perspective has attracted a lot of attention (see \cite{Buchert2008} for a recent review on this issue): it may even offer a possible solution to explain the acceleration of the universe, within the context of general relativity and without the introduction of any further exotic ingredient.
  T.~Buchert \cite{Buchert2000} derived the exact dynamical equations 
governing the average size $a_{\mathcal D}$ of a domain $\mathcal D$ in an inhomogeneous matter distribution. These equations can be written in the following form :
\begin{equation}
\left(\frac{\dot {a}_{\mathcal D} }{a_{\mathcal D}}\right)^2 = \frac{8\pi G\rho }{3}  
-\frac{{\mathcal Q}}{6}-\frac{{\mathcal R}}{6}\\
\label{eq:buchert}
\end{equation}
\begin{equation}
\left(\frac{\ddot {a}_{\mathcal D} }{a_{\mathcal D}}\right) = -\frac{4\pi G\rho }{3}  
+\frac{{\mathcal Q}}{3}
\label{eq:buchert2}
\end{equation}
for irrotational dust, and after dropping the cosmological constant for clarity.
In this expression, $\mathcal R$ is the average curvature over the domain and $\mathcal Q $  is an  average quantity over the domain $\mathcal D$:
\begin{equation}
 {\mathcal Q} = \frac{2}{3}\left(\langle \theta^2 \rangle-\langle \theta \rangle^2  -2\langle \sigma^2 \rangle \right)
\label{}
\end{equation}
$\theta$ being $3({\dot {a}_{\mathcal D} }/{a_{\mathcal D}})$ and $\sigma$ is the 
rate of shear $\displaystyle \sigma^2 = \frac{1}{2}\sigma_{ij}\sigma_{ij}$ ($\sigma_{ij}$ being the shear tensor).

 Given the analogy with the classical EFL 
equations, the possible consequence of back reaction has been discussed on this ground \cite{Buchertetal2000,BuchertCarfora2003,Buchert2008}.  

Following Buchert's work several authors regard the back reaction problem 
as an essentially non perturbative question which cannot be solved without an appropriate treatment \cite{Rasanen2008,Kolbetal2006}. 

 The question was handled from an other point of view by
looking at second order effect from superhorizon perturbations generated during inflation \cite{Barausse2005,Kolbetal2005}. The conclusion was  that some acceleration could be generated
in this way. However, this result has been disproved \cite{Flanagan2005,Geshnizjani2005,HirataSeljak2005,Giovannini2006,Rasanen2006}. 
The same conclusion holds for 
 superhorizon perturbations 
in presence of a scalar field \cite{KumarFlanagan2008}. The possibility remains open that back reaction from subhorizon scales induces a modification of the equations governing the expansion \cite{Rasanen2006,Kolbetal2006} and observational relations.
However, from the astrophysical point of view what matters precisely 
is whether  back reaction could alter significantly cosmological observations
and fools the traditional interpretation. Clearly if the back reaction has to 
modify cosmological quantities by a factor $h^2 \approx 10^{-10}$ the study of 
back reaction would remain of pure academic interest. Actually, modifications 
at the level of $h \approx 10^{-5} $ are expected from linear  perturbations at the scale of our horizon. Indeed, \cite{Wetterich2003} examined back reaction in a perturbative approach and concluded that gravitational energy of structure contributes as a back reaction term, but unless 
 dark matter is essentially made of black holes, no appreciable change beyond
the $10^{-5} $ level is to be anticipated. Such a term should be called a weak back reaction, because it can be  anticipated from simple dimensional arguments.
The important  question is~: are there strong back reaction terms, i.e. 
terms which arise from non linear effects and which could contribute at a level much larger than the naive order of magnitude $\approx 10^{-10}$ or even be of the order of  
being dynamically as important as the density $\rho$.  
At this level we are facing traditional difficulties when handling with relativistic effects. For instance, Buchert equation are written in a comoving frame
in which particles are at rest. Therefore the non-zero ``mass'' associated 
with the kinetic energy of particles should appear as a ``back reaction'' 
term in the EFL equations. Identically the energy associated to a gravitational field 
has to be taken into account.  These are weak back reaction terms. Using Newtonian perturbed RW metric \cite{nogo2006} 
showed  that back reaction term can entirely disappear from the  second EFL 
equation \ref{eq:EFL2} when the density is  properly evaluated, but with a different choice of coordinates, and concluded that
 we are very close to a ``no-go'' theorem: ``no cosmic acceleration occurs as a 
result of the nonlinear
back reaction via averaging'' (on this question a major source of complication arises
 from the gauge issue and different authors may find apparently discrepant 
results because of the use of different gauges). The present formulation of Buchert's equation is likely to depend explicitly on the choice of the coordinate
system and therefore the relevance to observations is unclear. It is not even clear whether a clever gauge transformation would not allow to suppress entirely this 
back reaction term. Similar conclusions were 
reached by 
\cite{NoGo2}, who emphasized again that gravitational energy is a source of 
the gravity field which can appear as a ``back reaction'' term, i.e. is a  weak back reaction term, and could be qualified as 
such, in agreement with \cite{Wetterich2003}, but which remains small in 
practice. In addition, they showed in some exact solutions the absence of observable ``back reaction''. In this respect,
the Swiss cheese-model \cite{SwissCheese} is an interesting exact solution in which the presence 
of inhomogeneities (that could even be non linear i.e. $h \sim 1$) does not lead to 
any unexpected  behavior  i.e. in which there is no strong back reaction
effect. From a second order perturbation analysis \cite{NoGo2} also argued that as long the Newtonian conditions are satisfied:
\begin{equation}
l \ll \frac{c}{H_0} {\rm \ and \  } v \ll c
 \label{eq:condition}
\end{equation}
``there always exist gauges where the metric differs from the
RW metric only at order $\langle h\rangle$''. Of course, when the size of the
 perturbations
 becomes comparable to the Hubble radius,  one expects significant alteration 
of the standard relations \cite{Kolbetal2008}. A more troubling paper has been put on
 the ARXIV, but remains unpublished which : \cite{NambuTanimoto} investigated 
the solution of Buchert' equations in a exact spherical solution (the LTB that 
is described in the next section), actually a  very interesting approach. They concluded that at some point ``the universe starts accelerated expansion'',
 a similar conclusion was drawn by \cite{Chuang2005,Mansouri2005,Moffat2006,Mansouri2006}, while 
it is clear that from spherical solution no physical acceleration is possible
thanks to Birkhoff's theorem and as it has been confirmed by \cite{Alnesetal2007,EnqvistMattsson2007}. This is a direct indication that acceleration could appear in some quantities but which is by no means related to the observed acceleration. The question has fallen in the
 arena of subtle general relativity issues, but it seems that there are serious 
claims that
back reaction cannot produce an  acceleration comparable to what is requested from the   observations. There are also 
convincing indications that no correction to standard EFL equations is expected 
above the amplitude of first order perturbation $10^{-5}$. There is therefore 
an important   ``cosmological no-go theorem''which remains by now only a 
conjecture that 
for the standard CDM fluctuations spectrum, there is no significant back reaction
and the classical description is sufficient. It is by now the duty of advocates 
of real back reaction effetc to demonstrate that interesting terms may exist
beyond standard known linear and non-linear contributions.
It remains important to examine whether weak back reaction could raise changes of the order of $10^{-3}$ or more, as such an effect would start to be of the order of the precision anticipated for the
 determination of the cosmological parameters from future experiments. 
The fact that equation (\ref{eq:buchert}) looks like standard one , i.e. first EFL equation,  does not at all mean that terms appearing in it should be identified with 
similar looking standard cosmological parameters. These equations 
are valid for a given special choice of coordinates and it is therefore far 
from trivial to identify which observable consequences would follow. 
Even the term $\frac{\dot {a}_{\mathcal D} }{a_{\mathcal D}}$
should not a priori be identified to the standard Hubble expansion rate
nor should $\ddot {a}_{\mathcal D}$ be used in comparison with to $-q_0$ . 
Strictly speaking when observers determine the Hubble constant $H_0$ they compare 
the observed linear relation between luminosity  distance and redshift from
some galaxy sample, which are both observable quantities, and $q_0$ is related to the 
leading second order term in $z$. Identically, when the cosmological density 
parameter (for 
instance) is determined by fitting the CMB $C_l$, they do not proceed with 
the measurement of the actual density of the present day  universe. A further problem is the fact that quantities entering eq. \ref{eq:buchert} and  eq. \ref{eq:buchert2}  are averaged over some volume with some characteristic scale. They  are therefore expected to vary with this scale.
Indeed, virulent criticisms have been raised up against the (strong) back reaction program, 
pointing out that the metric eq. \ref{eq:perturb}, with  Friedmann-Lema\^\i tre  
equations have been fully successful to describe and to predict master pieces 
of observational cosmology, therefore any alternative model should demonstrate 
its ability to reproduce these data as well with the same level of {\em concordance}   \cite{IshibashiWald2006}.

Claims have
 been made recently for the detection of observable effects that could be attributed
to  the back reaction \cite{LiSchwarz2007}. Some models have been built 
\cite{Mattsson2007,Wiltshire2007} which were compared, successfully,  to the main 
cosmological data \cite{Leithetal2008,Larena09}, in agreement with the ``low'' Hubble 
constant found by  \cite{Tammann2008}.  One can hope  that this  is
 the sign that the 
actual importance of back reaction will soon be clarified, although Cosmology is a field were there is a tradition for opponents to the standard paradigm not to 
resign easily \cite{Jayant2008}.

\subsection{Large scale void(s)}

After the discovery of an apparent acceleration from the SNIa Hubble diagram
it has been emphasized  that the Hubble diagram in its own was not 
a direct proof of an actual acceleration: even if the assumption that 
SNIa are standard candles is correct, an apparent acceleration could be 
obtained from the Hubble diagram within an inhomogeneous cosmological model 
\cite{Celerier2000} when interpreted with the presumption of an homogeneous 
world. The  possible existence of large scale structure up to 
200 Mpc suggests that this option was to be considered \cite{Tomita2001}.
The existence of very large scale structure  has been a subject of regular 
claims in  Cosmology \cite{sliceCFA,Einasto1997}.
Such a possibility was regarded as a potential major failure of the standard 
view, and would even support the idea of a fractal distribution of matter on large scales \cite{FournierAlbe,MandelbrotCRAS,Grujic2002}. The availability of very large scale surveys like the SDSS and the 2DF
has almost entirely closed this issue: although the visual impression from
galaxy surveys might be  that structures  occasionally exist on very large 
scales, they are not 
the signature of average (r.m.s.) fluctuations on large scales greater than what could
 be anticipated for let say a standard $\Lambda$CDM spectrum, when measured with appropriate tools like the correlation function or the power spectrum.
There is therefore no serious piece of data that suggests that the average level of fluctuations
on large scales (as long as they remain smaller than the Hubble scale) is much larger than anticipated from the standard picture.
The anomalous character of some rare high fluctuations is still possible \cite{coldspot,HuntSarkar} and could for instance be due to some non-Gaussian features of the primordial fluctuations, but the 
tendency of the human brain \cite{Peebles93} to identify insignificant patterns is 
a serious worry when dealing with these questions.

Given that observations tell us 
that the universe is isotropic around us, if 
some large scale inhomogeneity has to exist and be in agreement with present day observations it has to be nearly spherical and
we should occupy a very specific place in this universe: for instance it may happen that we are close 
to the center of a nearly spherical perturbation. 
Such a possibility might have some theoretical motivation \cite{Lindeetal1995}.
If we actually live inside a gigantic void, close to its center (to preserve the observed isotropy of the sky), this 
requires the abandonment of the Einstein cosmological principle
that postulates a homogeneous matter distribution on large scales, and 
 of the Copernican principle, that we are not in a special location in the universe.

 The homogeneity of the universe is directly observable in principle 
on scales much smaller than the Hubble radius, it is by no way obvious 
that it can actually be tested from observations on the largest scales 
(standard textbooks of relativity 
often mentioned that it cannot be actually tested, but this is assuming 
a restricted set of observable quantities). Detailed investigations of inhomogeneous models were carried out earlier, but not essentially  with the purpose to provide an alternative explanation of the apparent acceleration, see for instance \cite{BarrettClarkson} for a description of various class of inhomogeneous
cosmological  solutions and references to earlier works on this subject. 

 The exact solution of the 
spherical inhomogeneity in general relativity was first published by  Lema\^\i tre \cite{Lemaitre1933}. Tolman \cite{Tolman1934} and Bondi  \cite{Bondi}, aware of Lema\^\i tre's work published studies on the same solutions. These solutions have been extensively used to examine closely whether the apparent acceleration could actually be due to such an inhomogeneous spherical solution \cite{Tomita2001,Tomita2001b,Iguchi2002,Bolejko,ChungRomano2006,Garfinkle,Moffat2006,EnqvistMattsson2007}. More complex inhomogeneous world solutions to Einstein equations have also been 
proposed to reproduce the observed SNIa Hubble diagram \cite{Ishak2007}.
Interestingly enough, \cite{Biswasetal2007} found that some ``onion'' structure
 could leave no strong apparent effect. Qualitatively speaking, a void is a 
region with a typical size $\cal L$ where the density is lower than the average, in order for the 
initial singularity to happen at the same ``initial time'' in the past, the expansion rate in the region should be higher. This requirement is however not mandatory and it is possible to build isotropic inhomogeneous solutions with singularity starting at a different time \cite{celerierschneider98}. Usually the ``true'' universe is 
an Einstein de Sitter $\Omega_M= 1$ and the parameters inside the void are such that it looks like a low density universe universe. So that inside the void:
\begin{equation}
\tilde{\Omega}_M \sim 0.25 {\rm \ and \ } \tilde{H}_0  \sim 0.72
\end{equation}
while asymptotically (outside the void):
\begin{equation}
{\Omega}_M = 1 {\rm \ and \ } {H}_0 \sim 0.50
\end{equation}
the value of the Hubble constant being set by the ``initial time'' constraint.
Roughly speaking in such models the distant universe has a  Hubble constant lower
than the local one, which is interpreted as an acceleration.  The minimal size 
of the void to produce an apparent acceleration from supernovae at $z \sim 0.5$
is typically of the order 300$h^{-1}$Mpc, or slightly less, i.e. up to 
$z \sim 0.1$ \cite{Alexanderetal2007}. However, this kind of consideration is
 not sufficient to offer a satisfying alternative: the Hubble flow is known to 
be quite  smooth and in agreement with an uniform expansion. A significant
 change of the Hubble constant over a volume of size smaller than the horizon size will lead to an 
appreciable change of 
the apparent Hubble constant. Actually, such effect has been 
found \cite{Jha2007}, who notice  a Hubble constant 6.5\% smaller on scales 
greater than 75$h^{-1}$Mpc, a small number that might be due to a problem in 
the calibration of close supernovae \cite{Conleyetal2007}. No other significant departure from an uniform
Hubble constant has been recently reported, actually the Hubble diagram appears 
to be remarkably regular from a few Mpc up to redshift where cosmological correction are needed \cite{MadoreFreedman1998,Kowalskietal2008,Tammann2008}. Therefore, one can expect that voids witch size $\cal L$ is  significantly smaller  than the Hubble radius will not mimic adequately the present day Hubble diagram. Appreciable difference would be easy to check in the near future. However, if the voids are large, comparable to the size of the Hubble radius,  it is intuitive that 
a good match to the supernova data could be achieved. \cite{Vanderveldetal2006} showed that in order to match closely the apparent acceleration of the Hubble diagram,  a singularity at the origin should be present in the metric and other pathologies may exist. It is therefore far from being obvious that dust-filled LTB can reproduce the apparent acceleration in detail. Indeed, \cite{Cliftonetal2008} concluded, that if one keeps
 the constraint that the curvature has to remain smooth around the origin, the 
Hubble diagram from an LTB will be different from the one in $\Lambda$CDM
and could be distinguished thereby  with enough statistics at low redshit or intermediate redshift. 

 Convincing examples of successive models have been obtained by \cite{Enqvist2008,JuanVoid1}, who also provided a public code : \url{ http://www.phys.au.dk/$\sim$haugboel/software.shtml}. For their best LTB model 
differences in the Hubble diagram between $\Lambda$CDM and LTB  appear to be 
less than $0.05$m. Such differences are probably due to the choice of the 
analytical profile, and might therefore being reduced (or fitted to future data), even if this would be at the price of introducing further free parameters. 
The local Hubble constant is rather low ($H_0 \sim 60$ km/s/Mpc) but still acceptable.
It would be interesting to examine whether this behavior is generic or specific to their model.
However, 
 the confidence in the $\Lambda$CDM relies on other pieces of evidence
 and the question of whether the CMB and the BAO feature can be reproduced 
has been also incorporated \cite{Alnes2006,Alexanderetal2007,JuanVoid1}.  Because the 
outer cosmology is an Einstein de Sitter Universe with a small Hubble constant, an appropriate fit can be obtained relatively easily provided that  the angular distance to the CMB 
is matched appropriately.  These are rough evaluations, as a detailed formalism 
to deal with perturbations in a LTB model is lacking. However, this might soon
 become available \cite{Clarkson2007,Zibin2008}, and it is reasonable to 
believe that an adequate fit can be obtained. It is nevertheless fair to say 
that these models are not yet as impressively good to fit cosmological data
as the concordance is,  perhaps because of their intrinsic complexity (but on 
the other hand, they contain many more degrees of freedom). The situation looks
 like that this type of hypotheses (large void) could be adjusted to the data 
and could not therefore be rejected from observations if enough tuning is 
allowed. However, fortunately, this is not the case! \cite{Uzan2008} have shown
 that in a LTB model,  the time drift of the cosmological redshift can be 
different from what it is in homogeneous worlds, providing a possible test of the
 Copernican principle.
The kinematic Sunyaev-Zel'dovich (kSZ)
effect on distant clusters will be significantly different for distant clusters offering a different method of testing LTB \cite{JuanVoid2}. Another impressive tool
for constraining LTB models has been proposed by \cite{CaldwellStebbins2008}: 
we know that the intergalactic medium which probably contains most of the 
baryon is  highly ionized up to redshifts  greater than 5; CMB photons are scattered by 
the electrons of this plasma, so photons we collect from the CMB 
are a mixture of photons which have travelled straight towards us and of photons which have been scattered at lower redshifts, if these electrons are 
within the void region, they are scattering a very different CMB because they are 
moving rapidly. In addition, even electrons which are out of the void will see a different CMB sky because the void 
itself and will be a further source of a distorted CMB. Therefore contrary to the homogeneous case,   an 
observer in the center will observe a combination of black-bodies with 
different temperatures, resulting in a distorted spectrum. Present day
 limits on possible spectral distortion of the CMB already provide interesting 
constraints and could become critically more stringent if limits could be improved 
by an order of magnitude \cite{CaldwellStebbins2008}. \\

\subsection{Quintessence}

A true cosmological constant is an object that is largely unwanted from the theoretical point of view. Ratra and Peebles introduced the idea that
the acceleration could be due to the presence of some scalar field dominating present day density of the Universe \cite{RatraPeebles1988,PeeblesRatra1988}. 
Quintessence names a scalar field that coupled to gravity.  The canonical Lagrangian 
of a scalar field is :
 \begin{equation}
{\cal L}  = \frac{1}{2}(\partial_i \phi)^2 -V(\phi)
\label{eq:Lquin}
\end{equation}
the first term being the kinetic energy ($X= \frac{1}{2}(\partial_i \phi)^2$), and the second one the potential. 
 The stress-energy tensor has a form identical to that of an ideal fluid with pressure and density given by~: 
\begin{equation}
p = {\cal L} = \frac{1}{2}(\partial_i \phi)^2 -V(\phi)
\label{}
\end{equation}
\begin{equation}
\rho = 2\ X  \partial_X {\cal L} -  {\cal L} = \frac{1}{2}(\partial_i \phi)^2 +V(\phi)
\label{}
\end{equation}
For a field which is spatially homogeneous the equation of state parameter $w$ is:
\begin{equation}
w = \frac{p}{\rho} =  \frac{\frac{1}{2}\dot{\phi}^2 -V(\phi)}{\frac{1}{2}\dot{\phi}^2 +V(\phi)}
\label{eq:wx}
\end{equation}
which remains $\geq -1$. The equation driving the field in a FL world is :
 \begin{equation}
\ddot{\phi}+3H\dot{\phi} = -\frac{dV}{d\phi}(\phi)
\label{eq:vphi}
\end{equation}
The original scenario was proposed \cite{RatraPeebles1988} with a potential of the form:
\begin{equation}
 V(\phi) = \frac{M^{4+\alpha}}{\phi^\alpha}
\label{}
\end{equation}
Let us suppose that we are at some late time when radiation or matter  dominates. Then
the expansion factor is $a(t) \propto t^n$ with $\displaystyle n = \frac{1}{2}, \frac{2}{3}$. A power law solution of equation (\ref{eq:vphi}) $\phi \propto t^\beta$ 
behaves as:
\begin{equation}
 \phi \propto t^{\frac{2}{\alpha+2}}
\label{eq:solattractor}
\end{equation}
and the ratio of the density of the scalar field to the total density is:
\begin{equation}
\frac{\rho_\phi}{\rho} \propto a^{\frac{4}{n(\alpha+2)}}
\label{}
\end{equation}
So that if $\alpha >0$ the density of the scalar field will be dominant. 
Another important property is that solution \ref{eq:solattractor} is an 
attractor in that it  describes the late solution of a large class of initial 
conditions. If the kinetic term becomes small at a later epoch in \ref{eq:wx}, 
the field behaves like a cosmological constant. This does not necessary solve 
the coincidence problem, but at least alleviates it. 

 Dark energy has become the subject of intense theoretical efforts,
which lead to investigate different forms of the Lagrangian. 
One first generalisation consists in modifying the potential $V(\phi)$. 
A more radical approach is to modify the kinetic term. 
In $k-$essence models \cite{Kess,Kess2}, inspired from $k-$inflation \cite{kinflation}, the Lagrangian is written in a general form  :
\begin{equation}
{\cal L} = p(\phi,X) 
\label{}
\end{equation}
and 
 contains therefore a non canonical kinetic term.

Few constraints can be added. It can be requested that the density $\rho $ remains positive, although there is no decisive argument against a negative cosmological constant and so this constraint might not be necessary. On the other hand 
such a term would lead to a rapid big crunch, which is unwanted.
Another possible constraint is that  the square of the sound's speed remains positive \cite{kinflation}:
\begin{equation}
 c_s^2 = \frac{\partial_{\small X} p}{\partial_{\small X} \rho} =  \frac{{\cal L}_{,\small X}}{{\cal L}_{,\small X}+2X{\cal L}_{,\small XX}}
\label{}
\end{equation}
($,\small X$ stands for $\displaystyle \frac{\partial}{\partial X}$ and $,\small XX$ $\displaystyle \frac{\partial^2}{\partial X^2}$).

Such models easily lead to large variations of   $c_s$ (while quintessence based on  (\ref{eq:Lquin}) automatically leads to $c_s  = 1$) and one can have $c_s  >1$, 
this is not problematic because the theory remains Lorentz invariant, so there is no violation of causality \cite{Kess3,Babichevetal2008}\footnote{ For a different point of view see \cite{Bonvin2006,Ellisetal2007}.}. Within such models  the equation of state parameter $w$ could take any value (\cite{Melchiorrietal2003}) and so one can have:
\begin{equation}
w < -1.
\label{ew:w<-1}
\end{equation}
Such dark energy is referred to as  phantom (ghost) energy. A simple example is obtained by changing the sign of the kinetic energy term in \ref{eq:Lquin}. \\

So from the astrophysical point of view $w$ can be regarded as a free function which values  should be determined from observations. From the fundamental point of view $V(\phi)$ or the Lagrangian itself  are the quantities to be determined and therefore observations have to be used to constrain directly these 
quantities, rather than the parameter $w(z)$. Finally, let us mention that more 
details on scalar field models of dark energy can be found in the recent review
\cite{Copeland2006}.

\subsection{New gravity law on large scales}

The existence of a large scale acceleration in the Universe is a serious indication  that our present knowledge of gravity is actually incomplete. Although the introduction of a new component, quintessence, might provide the explanation for  the acceleration, i.e. a new term in the energy momentum tensor, the historical cosmological constant introduced a new term in the other side of Einstein equation. Therefore it might well be that the geometrical term is more complex that we used to believe. This option of modifying gravity to account for the acceleration is briefly introduced in this section. More details can be found in recent reviews on this topic \cite{revRDRM,revDEMG}.\\

While  General Relativity can be derived from the Einstein-Hilbert action:
\begin{equation}
{\cal S} = \int dx^4 R 
\label{eq:EH}
\end{equation}
where $R$ is the Ricci scalar, one can try a more general action, leading to the so called $f(R)$ theories:
\begin{equation}
{\cal S} = \int dx^4 f(R) 
\label{eq:fR}
\end{equation}
where $f$ is an unknown function. Such theories have to pass solar system 
tests as well as pulsar chronometry tests  on small scales and cosmological 
tests on large scales. In practice such theories met considerable problems, 
and it is far from being clear that viable models can be built in this way. An 
alternative procedure, is the so called Palatini formalism, in which the 
variation principle is modified: the metric and the connection are regarded as independent quantities, and the resulting equations are in general different but for $f(R)\propto R$.  Within this formalism, \cite{AmarzguiouiEtal}
have examined constraints on a specific model:
\begin{equation}
 f(R)  = R \left(1+ \alpha\left(\frac{-R}{H_0^2}\right)^{\beta-1} \right)
\label{eq:fR2}
\end{equation}
where the second term is essentially a correction term to the classical Einstein-Hilbert Lagrangian.
Acceptable regions of the parameter space encompass the standard $\Lambda$CDM
model, but these models differ from standard cosmological models, as  the relations between $\Omega_k$, $\Omega_m$, $q_0$ and the growth rate $\displaystyle \frac{ d\ln D}{d\ln a}$
are different. 

A more radical way to modify gravity in our world is through the idea of higher
 dimensional space. Such propositions can be traced back to 1919  with the 
work of Kaluza \cite{Kaluza} who proposed an unification scheme for gravity
 and electromagnetism within a fifth dimensional space. \cite{Klein} pointed 
out the interest of having a fifth compact dimension  to avoid observational 
constraints on large additional dimension. Superstring theory and supergravity theories possess remarkable 
properties in higher dimensional space and have therefore deserved strong attention from theorists. Modern versions are known as braneworlds or brane cosmology
\cite{brane1,Bin00a}.
Our 3+1 world lies on the brane, while the remaining space is the bulk. Matter, with pressure and density, 
is present only in the brane, but gravity is present in all dimensions.
The vacuum energy in the brane provides a tension term $\sigma$ and there is an other vacuum energy in the bulk $\Lambda_B$.
The general action now contains terms involving an equivalent of the Ricci scalar  corresponding to gravity in higher dimension. The inferred equations describing the expansion on the brane, the generalized FL equations, contain more new  terms which could be non-linear and which are related to the properties of the bulk \cite{Bin00b}: 
\begin{equation}
H^2   \propto  \cdots \rho^2  + \cdots \Lambda_B + \frac{\cal C}{a(t)^4} - \frac{k}{a(t)} 
\label{eq:brane}
\end{equation}
 the third additional term behaves like radiation but comes from the bulk and is sometime called the dark radiation on which BBN provides stringent constraints
\cite{Bin00a,Bin00b,Flanaganetal2000,Iocco08}.

Direct modification of the FL equations have also been proposed through Cardassian models \cite{cardassian}:
\begin{equation}
H^2   = A\rho+B\rho^n 
\label{eq:cardassian}
\end{equation}
which does not contain a vacuum contribution, although in models  with $n=0$ the second term of the right hand side behaves identically to a $\Lambda$  term. Such a form for the  EFL equation can arise from some 
particular fluid properties of dark matter or from higher dimensions \cite{GondoloFreese}. 

In DGP models, extra dimensions are infinite \cite{DGP}, and the effective 
action contains explicitly a 4-D Einstein-Hilbert term i.e. on the brane in addition to the 5-D term on the bulk. This introduces a scale $r_c$ and two distinct regimes
appear: on scales smaller than $r_c$  gravity essentially results from the 4-D term and classical GR is recovered (although high precision tests might lead to some differences \cite{dgp:solarsystem,dgp:solarsystem2}); on larger scales the gravity is ``leaking'' the expansion is eventually accelerated. The FL equation is replaced by (\cite{Deffayet2001}):
\begin{equation}
H^2   = \left(\sqrt{\frac{\rho}{3M^2_{Pl}}+\frac{1}{4r^2_{c}}}+\epsilon\frac{1}{2r_c}\right)^2
\label{eq:brane2}
\end{equation}
with $\epsilon = \pm 1$. From this it is clear that at early times $\displaystyle 
\frac{\rho}{3M^2_{Pl}} \gg \frac{1}{4r^2_{c}}$ allows to recover the classical 
EFL regime, while at late times the  accelerated expansion is recovered (provided that  
$\epsilon =+1$).  Detailed predictions of DGP models may not be identical to those of 
$\Lambda$CDM, and recent investigations found some tension between data and 
theory \cite{dgp:obs,Rubinetal2008,Fangetal2008}, although the predictions are not as 
straightforward as in the standard model. From the astrophysical point of view, 
an interesting aspect of these classes of models in which gravity is modified is 
that relations between cosmological parameters are not identical to those in the $\Lambda$CDM and the evolution of the growth factor is expected to be 
different \cite{Linder2005}, with a possible dependence on scales, allowing for possible discrimination between various possible origins for the acceleration \cite{Uzan,Bertschinger2008}. 

A more radical option has emerged in recent years~: given the fact that the accelerated expansion clearly needs  some revision of our knowledge of the gravitational sector, could it be that the problem of dark matter 
itself is due to the break of standard gravity laws at finite distance? This possibility has been introduced 
in an empiric way by Milgrom  in order to explain the rotation curves of galaxies without invoking dark matter \cite{Milgrom}. A fully relativistic theory has been built by Bekenstein \cite{Bekenstein} which 
behaves like Milgrom' law in the weak field regime: the TeVeS (tensor-vector-theory) theory. Quite remarkably,
it has been shown recently that this theory can reproduce the observed cosmic acceleration,  large scale power spectrum (at the time of the work) and  CMB 
 with comparable success as  $\Lambda$CDM  \cite{teves:obs}. 
The present status of this theoy is not as satisfactory as with the standard concordance model~:   this theory did not lead to specific predictions which  have been verified a posteriori, and
 it needs the introduction of a  new type of fields in physics in the form of vector fields. However, MOND is a clear example of an alternative view, which differs drastically in its fundamental ingredients and 
illustrates  the fact that our understanding of the gravitation sector relevant to cosmology might be more limited than commonly assumed.

\section{The area of precision Cosmology}

\begin{table*}[ht!]

\centering
\begin{tabular}{c c c c c }
\hline
$\mathbf{Parameter}$		&$\mathbf{Vanilla}$	 &$\mathbf{Vanilla + \Omega_k}$			&$\mathbf{Vanilla+w}$ & $\mathbf{Vanilla+\Omega_k+w}$   \\
\hline
$        \Omega_b h^2$      &$ 0.0227\pm 0.0005$                            &   $0.0227\pm0.0006$           & $0.0228\pm0.0006$      & $0.0227\pm0.0005$ 	         \\
$           \Omega_c h^2$      &$0.112\pm0.003 $                     &  $0.109\pm0.005$            & $0.109\pm0.005$      & $0.109\pm0.005$	        \\
$                 \theta$      &$ 1.042\pm0.003$                       &  $1.042\pm0.003$            & $1.042\pm0.003$	    & $1.042\pm0.003$	 	\\
$                 \tau$      &$ 0.085\pm0.017$                       &  $0.088\pm0.017$            & $0.087\pm0.017$	    & $0.088\pm0.017$	        \\
$             n_s$      &$0.963\pm0.012$                             &  $0.964\pm0.013$            & $0.967\pm0.014$	    & $0.964\pm0.014$	        \\
$                       \Omega_k$      &$ 0$                         &  $-0.005\pm0.007$            &  $ 0 $	    & $-0.005\pm0.0121$	         \\
$                        w$      &$ -1 $                             &  $ -1$            &  $-0.965\pm0.056$	    & $-1.003\pm0.102$	        \\
\hline
$                      \Omega_{\lambda}$      &$ 0.738\pm0.015                $   &  $0.735\pm0.016$           & $0.739\pm0.014$      & $0.733\pm0.020$		\\
$                      Age       $      &$  13.7\pm0.1$              &  $13.9\pm0.4$            & $13.7\pm0.1$	    & $13.9\pm0.6$	        \\
$                      \Omega_M$      &$ 0.262\pm0.015$                   &  $0.270\pm0.019$            & $0.261\pm0.020$	    & $0.272\pm0.029$		\\
$                    \sigma_8$      &$0.806\pm0.023$                 &  $0.791\pm0.030$            & $0.816\pm0.014$      & $0.788\pm0.042$		 \\
$                   z_{re}$      &$10.9\pm1.4$                       &  $11.0\pm1.5$            & $11.0\pm1.5$	    & $11.0\pm1.4$		\\
$                    h $      &$ 0.716\pm0.014$                      &  $0.699\pm0.028$            & $0.713\pm0.015$	    & $0.698\pm0.037$		\\

\hline
\end{tabular}
\label{tab:cosmopar}
\caption{Summary  of the  mean values and  68\% confidence intervals for the  parameters constrained from CMB, SNIa and BAO for different models ($ \theta$ is the ratio of sound horizon to angular diameter distance). These constraints are quite tight, most of them are below 5\%, and are stable  when  additional degrees of freedom are added to the model ($w$,$\Omega_k$), adapted from \cite{lfabyz}. 
}
\end{table*}                

After the discovery of the fluctuations of the  cosmological background by COBE
the perspective to achieve precision measurements of the angular power spectrum
of these fluctuations  appears to be within reach and two satellite experiments were 
designed to reach this target: WMAP and Planck. This has opened a new avenue 
for Cosmology to benefit from high precision measurements with well controlled 
systematics. Indeed, WMAP delivered data of high accuracy, allowing for high precision estimations of the cosmological parameters \cite{spergel2003,dunkley}. Furthermore, the consistency  of SNIa, BAO and  CMB data allows reliable accurate estimations by combining constraints \cite{Komatsu,Kowalskietal2008}.
 The possibility of reaching high precision measurements with
 different techniques (distant SNIa with SNAP, properties of distant x-ray Clusters  as proposed by 
 Panoramix some years ago or the recent WFXT \cite{WFXT}, or with the SZ sample of clusters expected with Planck and other CMB experiment like SPT,  full 
sky weak lensing surveys with DUNE, or the more recently proposed redshift surveys 
of hundred millions of galaxies with SPACE or ADEPT) has not only reinforced 
the perspective to determine  cosmological parameters with  high precision, 
but will also allow to investigate the very nature of  dark energy. The Dark Energy 
Task Force report \cite{DETF} presented a summary of the different levels of 
progress expected for various  projects. Ultimate constraints will be 
obtained by combination of various experiments. It is interesting to summarize 
what present day data allow. There are many possible data to combine, but the 
three most currently used are SNIa, CMB and measure of the large scale 
distribution, either through the power spectrum or through the  correlation function. 
A combination of these three data sets leads to tight constraints on the minimal
 $\Lambda$ Cold Dark Matter model. These constraints are summarized in Table 
\ref{tab:cosmopar} adapted from \cite{lfabyz}. Different groups recently got similar results with  approaches which slightly differ in the technical details \cite{Komatsu,Kowalskietal2008,Sanchez}, so these numbers can be regarded as very robust. As one can see, most of the cosmological parameters describing our Friedmann-Lema\^\i tre universe are constrained
 to an accuracy better than 5\%. Furthermore, when the parameter space is enlarged the constraints remain essentially unchanged.
This calls for some caution. Quoted uncertainties
reflects statistical uncertainties.  Unidentified systematics are the critical 
issues in this topic. Variations in published values of $\sigma_8$ from various 
approaches (CMB, Clusters, weak lensing) have provided an illustration that 
systematics uncertainties could alter preferred values beyond statistical 
uncertainties. On the other hand,  given possible systematics which have been identified until now, it is likely that future 
estimations
will remain within the two sigma domains. First, when combining only 
two probes, one already gets tight constraints which are within this range. A 
second argument is that when one allows for additional parameters (free 
equation of state parameters $w$, curvature, non-zero neutrino mass, tensor 
contribution, ...) preferred values and interval ranges are not changed by 
much. These are indications that we are already in the precision area of cosmology: present day estimations of cosmologcail parameters are likely not to change by much in the near future and investigations of the nature of dark energy will need
extremely accurate control of systematics. Whether the necessary investments
will be valuable for the astronomical field is a subject of debate \cite{White07}.
     
\section{Conclusions}

The Copernician model of the world  was the first revolution of a series in the 
construction of modern cosmology, and the discovery of the accelerated 
expansion being the latest in date. Theoretical considerations have always been
a source of remarkable observational investigation and   
Cosmology has always benefited from the confrontation of models with 
observations. Since the thirties, the big bang picture, the modern version of 
Lema\^\i tre's primeval atom has been remarkably successful, based on simple 
assumptions and physics laws that have been validated  by accurate experimental 
results. Although alternative theories have been developed, these alternative
were based on hypothetical unknown physics advocated to interpret cosmological 
observations. None of these alternative theories has produced predictions that 
have been comforted a posteriori. Rather new observations in agreement with
predictions of the big bang picture necessitated deep revision of the  
unorthodox view, at the cost of rather ad hoc assumptions added to fit the
new observations. The situation has evolved when the standard picture has 
necessitated the introduction of new ingredients, first dark matter and more 
recently dark energy. The very nature of these new ingredients, which are 
supposed to dominate the mean density of the universe has not been established 
by direct laboratory experiments, nor by astronomical observations, and this situation may some time lead to the 
question whether cosmologists have not  introduced new aethers.
We had the opportunity to see that the situation is not so. The introduction of 
-cold- non-baryonic dark matter has led to specific predictions, the amplitude of 
the fluctuations of the cosmological background on various angular scales, which were verified with high 
accuracy precision. The presence of dark energy has lead to a specific prediction,
the shape of the power spectrum on large scales, which has been verified  a posteriori. Although, the 
inclusion of a cosmological constant was concomitant to general relativity, 
the actual origin of dark energy remains totally unknown and the presence
 of dark energy in the present day universe represents probably the 
most fundamental and unexpected new element in modern physics.
  
  



\begin{thebibliography}{1}
%
%
\bibitem{abra} Abramowicz, M.~A., 
Bajtlik, S., Lasota, J.-P., \& Moudens, A.\ Eppur si Espande, Acta Astronomica, 57, 139 (2007)
\bibitem{DETF} Albrecht, A., et al.\  Dark Energy Task Force report to the Astronomy and Astrophysics Advisory Committee,
\url{http://www.nsf.gov/mps/ast/detf.jsp}, arXiv:astro-ph/0609591 (2006)
\bibitem{Alexanderetal2007} Alexander, S., 
Biswas, T., Notari, A., 
\& Vaid, D., Local Void vs Dark Energy: Confrontation with WMAP and Type Ia Supernovae, ArXiv e-prints, 712, arXiv:0712.0370 (2007)
\bibitem{Allen2008} Allen, S.~W., Rapetti, 
D.~A., Schmidt, R.~W., Ebeling, H., Morris, R.~G., 
\& Fabian, A.~C.\ Improved constraints on dark energy from Chandra X-ray observations of the largest relaxed galaxy clusters, \mnras, 383, 879 (2008)
\bibitem{Alnesetal2007} Alnes, H., Amarzguioui, 
M., \& Gr{\o}n, {\O}., Can a dust dominated universe have accelerated expansion?, Journal of Cosmology and Astro-Particle Physics, 1, 7 (2007) 
\bibitem{Alnes2006} Alnes, H., Amarzguioui, 
M., \& Gr{\o}n, {\O}., Inhomogeneous alternative to dark energy?, \prd, 73, 083519 (2006)
\bibitem{AmarzguiouiEtal} Amarzguioui, M., Elgar{\o}y, {\O}., Mota, D.~F., \& Multam{\"a}ki, T.,Cosmological constraints on f(R) gravity theories within the Palatini approach, \aap, 454, 707 (2006)
\bibitem{brane1} Arkani-Hamed, N., 
Dimopoulos, S., 
\& Dvali, G., The Hierarchy Problem and New Dimensions at a Millimeter, Physics Letters B, 429, 263  (1998)
ArXiv High Energy Physics - Phenomenology e-prints, arXiv:hep-ph/9803315 
\bibitem{kinflation} 
Armendariz-Picon, C., Damour, T., 
\& Mukhanov, V., k-inflation, Physics Letters B, 458, 209 (1999)
\bibitem{Kess} 
Armendariz-Picon, C., Mukhanov, V., 
\& Steinhardt, P.~J.\ Dynamical Solution to the Problem of a Small Cosmological Constant and Late-Time Cosmic Acceleration, Physical Review Letters, 85, 4438 (2000)
\bibitem{Kess2} Armendariz-Picon, C., Mukhanov, V., 
\& Steinhardt, P.~J., Essentials of k-essence, \prd, 63, 103510 (2001)
\bibitem{Babichevetal2008} Babichev, E., 
Mukhanov, V., \& Vikman, A., k-Essence, superluminal propagation, causality and emergent geometry, Journal of High Energy Physics, 2, 101 (2008)
\bibitem{Barausse2005} Barausse, E., 
Matarrese, S., \& Riotto, A., Effect of inhomogeneities on the luminosity distance-redshift relation: Is dark energy necessary in a perturbed universe?, \prd, 71, 063537 (2005)
\bibitem{Bardeen1986} Bardeen, J.~M., Bond, 
J.~R., Kaiser, N., \& Szalay, A.~S.\ The statistics of peaks of Gaussian random fields, \apj, 304, 15 (1986)
\bibitem{BarrettClarkson} Barrett, R.~K., \& Clarkson, C.~A., Undermining the cosmological principle: almost isotropic observations in inhomogeneous cosmologies, Classical and Quantum Gravity, 17, 5047 (2000)
\bibitem{OP} Bartlett, J.   et al. \ The XMM-Newton Omega Project,  proceedings of the XXIth rencontres de Moriond, 
astro-ph/0106098  (2001)
\bibitem{dgp:solarsystem2} Battat, J.~B.~R., 
Stubbs, C.~W., \& Chandler, J.~F., Solar system constraints on the Dvali-Gabadadze-Porrati braneworld theory of gravity, \prd, 78, 022003 (2008)
\bibitem{Beck&M} Beck, C., \& Mackey, M.~C., Could dark energy be measured in the lab?, Physics Letters B, 605, 295 (2005)
\bibitem{Bekenstein} Bekenstein, J.~D., Relativistic gravitation theory for the modified Newtonian dynamics paradigm, 
\prd, 70, 083509 (2004)
\bibitem{Archeops} Beno{\^i}t, A., et al.\ The cosmic microwave background anisotropy power spectrum measured by Archeops, \aap, 399, L19 (2003)
\bibitem{ArcheopsPar} Beno{\^i}t, A., et al.\ Cosmological constraints from Archeops, \aap, 399, L25 (2003)
\bibitem{Boomerang} de Bernardis, P., 
et al.\ A flat Universe from high-resolution maps of the cosmic microwave background radiation, \nat, 404, 955 (2000)
\bibitem{Bertschinger2008} Bertschinger, E., \& Zukin, P., Distinguishing modified gravity from dark energy, \prd, 78, 024015 (2008)
\bibitem{Bin00a} Bin\'etruy, P., 
Deffayet, C., \& Langlois, D., Non-conventional cosmology from a brane universe., Nuclear Physics B, 565, 269 (2000)
\bibitem{Bin00b} Bin{\'e}truy, P., 
Deffayet, C., Ellwanger, U., 
\& Langlois, D., Brane cosmological evolution in a bulk with cosmological constant, Physics Letters B, 477, 285 (2000)
\bibitem{Birkhoff} Birkhoff, G. D.,  Relativity and Modern Physics, Harvard University Press (1923)
\bibitem{Biswasetal2007} Biswas, T., Mansouri, R., \& Notari, A., Non-linear 
structure formation and 'apparent' acceleration: an investigation, Journal 
of Cosmology and Astro-Particle Physics, 12, 17 (2007)
\bibitem{BiswasNotari2008} Biswas, T., \& Notari, A., 'Swiss-cheese' inhomogeneous cosmology and the dark energy problem, Journal of Cosmology and Astro-Particle Physics, 6, 21 (2008)
\bibitem{bvm} Blanchard, A., Valls-Gabaud, D., \& Mamon, G.~A.\ The origin of the galaxy luminosity function and the thermal evolution of the intergalactic medium, \aap, 264, 365 (1992)
\bibitem{BB} Blanchard,  A., Bartlett,  J. \  What does cluster redshift evolution reveal?,  \aap,  {332}, 49L (1998) 
\bibitem{B2000} Blanchard, A., Sadat, R., Bartlett, J.~G., \& Le Dour, M.\ A new local temperature distribution function for X-ray clusters: cosmological applications, \aap, 362, 809 (2000)
\bibitem{ATM} Blanchard, A., \& Douspis, M.\ Evidence for new physics from clusters?, \aap, 436, 411 (2005)
\bibitem{2003A&A...412...35B} Blanchard,  A., Douspis, M., 
Rowan-Robinson, M., \& Sarkar, S. \   An alternative to the cosmological ``concordance model'', \aap  {412}, 35 (2003)
\bibitem{2006A&A...449..925B} Blanchard, A., Douspis, M., Rowan-Robinson, M., \& Sarkar, S.\ Large-scale galaxy correlations as a test for dark energy, \aap, 449, 925 (2006)
\bibitem{BSBL} Blanchard, A., Sadat,  R.,  Bartlett., J.,  Le Dour, M., A new local temperature distribution function for X-ray clusters: cosmological applications  \aap  {362}, 809 (2000)
\bibitem{Bolejko} Bolejko, K., Supernovae Ia observations in the Lema\^\i tre--Tolman model, PMC Physics A, 2, 1 (2008)
\bibitem{Bonvin2006} Bonvin, C., Caprini, C., 
\& Durrer, R., No-Go Theorem for k-Essence Dark Energy, Physical Review Letters, 97, 081303 (2006)
\bibitem{Borg99} Borgani, S., Rosati, 
P., Tozzi, P., \& Norman, C.\ Cosmological Constraints from the ROSAT Deep Cluster Survey, \apj, 517, 40 (1999)
\bibitem{Branchi1} Branchina, V., Di 
Liberto, M., 
\& Lodato, I.\ Dark energy and Josephson junctions, Journal of Cosmology and Astro-Particle Physics, 8, 11 (2009)
\bibitem{Branchi2} Branchina, V., \& Zappal{\`a}, D., Time evolution of $T_{\mu\nu}$  and the cosmological constant problem, General Relativity and Gravitation, 42, 141 (2010)
\bibitem{Brandenberger2002} Brandenberger, R.~H., Back Reaction of Cosmological Perturbations and the Cosmological Constant Problem,  
 ArXiv High Energy Physics - Theory e-prints, arXiv:hep-th/0210165 (2002)
\bibitem{Bondi} Bondi, H., Spherically symmetrical models in general relativity, \mnras, 107, 410 (1947)
\bibitem{Casimir} Bressi, G., Carugno, G., 
Onofrio, R., \& Ruoso, G., Measurement of the Casimir Force between Parallel Metallic Surfaces, Physical Review Letters, 88, 041804 (2002)
\bibitem{Buchert2000} Buchert, T., On Average Properties of Inhomogeneous Fluids in General Relativity: Dust Cosmologies, General Relativity and Gravitation, 32, 105 (2000)
\bibitem{Buchertetal2000} Buchert, T., Kerscher, 
M., \& Sicka, C., Backreaction of inhomogeneities on the expansion: The evolution of cosmological parameters, \prd, 62, 043525 (2000)
\bibitem{BuchertCarfora2003} Buchert, T., \& Carfora, M., Cosmological Parameters Are Dressed, Physical Review Letters, 90, 031101 (2003)
\bibitem{Buchert2008} Buchert, T., Dark Energy from structure: a status report, General 
Relativity and Gravitation, 40, 467 (2008)
\bibitem{CaldwellStebbins2008} Caldwell, R.~R., \& Stebbins, A., A Test of the Copernican Principle, Physical Review Letters, 100, 191302 (2008)
\bibitem{celerierschneider98}C{\'e}l{\'e}rier, M.-N., \& Schneider, J., A solution to the horizon problem: a delayed big bang singularity, Physics Letters A, 249, 37 (1998)
\bibitem{Celerier2000} C{\'e}l{\'e}rier, M.-N., Do we really see a cosmological constant in the supernovae data?, \aap, 353, 63 (2000)
\bibitem{Chodo2007}  Chodorowski M.J, (2007) A direct consequence of the Expansion of Space? MNRAS 378 239-244
\bibitem{ChungRomano2006} Chung, D.~J.~H., \& Romano, A.~E., Mapping luminosity-redshift relationship to Lemaitre-Tolman-Bondi cosmology, \prd, 74, 103507 (2006) 
\bibitem{Chuang2005} Chuang, C.-H., Gu, J.-A., \& Hwang, W.~P., Inhomogeneity-Induced Cosmic Acceleration in a Dust Universe,
Classical and Quantum Gravity, 25, 175001 (2008) 
\bibitem{Clarkson2007} Clarkson, C., Covariant approach for perturbations of rotationally symmetric spacetimes, \prd, 76, 
104034 (2007)
\bibitem{Cliftonetal2008}  Clifton, T., Ferreira, 
P.~G., \& Land, K.,Living in a Void: Testing the Copernican Principle with Distant Supernovae, ArXiv e-prints,  Physical Review Letters, 101, 131302 (2008)
\bibitem{Conleyetal2007} Conley, A., Carlberg, R.~G., Guy, J., Howell, D.~A., Jha, S., Riess, A.~G., \& Sullivan, M., Is There Evidence for a Hubble Bubble? The Nature of Type Ia Supernova Colors and Dust in External Galaxies, \apjl, 664, L13 (2007)
\bibitem{cook} Cook, R.~J., \& Burns, M.~S.\ Interpretation of the cosmological metric, American Journal of Physics, 77, 59 (2009)
\bibitem{Copeland2006} Copeland, E. J., Sami, M., \& Tsujikawa, S., Dynamics of dark energy, Int. J. Mod. Phys. D15, 1753  (2006) 
\bibitem{Crocce} Crocce, M., Fosalba, P., 
Castander, F.~J., \& Gazta{\~n}aga, E.\ Simulating the Universe with MICE: The abundance of massive clusters, \mnras, 403, 1353 (2010)
\bibitem{sliceCFA} de Lapparent, V., 
Geller, M.~J., \& Huchra, J.~P., A slice of the universe, \apjl, 302, L1 (1986)
\bibitem{Deffayet2001} Deffayet, C., Cosmology on a brane in Minkowski bulk, Physics 
Letters B, 502, 199 (2001)
\bibitem{revRDRM} Durrer, R., \& Maartens, R., Dark energy and dark gravity: theory overview, General Relativity and Gravitation, 40, 301 (2008)
\bibitem{Durreretal2003} Durrer, R., Gabrielli, 
A., Joyce, M., \& Sylos Labini, F.\ Bias and the Power Spectrum beyond the Turnover, \apjl, 585, L1 (2003) 
\bibitem{DGP} Dvali, G., Gabadadze, G., 
\& Porrati, M., Metastable gravitons and infinite volume extra dimensions, Physics Letters B, 484, 112 (2000)
\bibitem{dunkley} Dunkley, J., et al.\ 
Five-Year Wilkinson Microwave Anisotropy Probe Observations: Likelihoods and Parameters from the WMAP Data, \apjs, 180, 306 (2009)
\bibitem{DyerRoeder} Dyer, C.~C., \& Roeder, R.~C., The Distance-Redshift Relation for Universes with no Intergalactic Medium, \apjl, 174, L115 (1972) 
\bibitem{Ebeling07} Ebeling, H., Barrett, 
E., Donovan, D., Ma, C.-J., Edge, A.~C., 
\& van Speybroeck, L.\ A Complete Sample of 12 Very X-Ray Luminous Galaxy Clusters at $z >$ 0.5, \apjl, 661, L33 (2007)
\bibitem{Eddington1930} Eddington, A.~S., On the instability of Einstein's spherical world, \mnras, 90, 668 (1930)
\bibitem{MassFunctionEfstathiou} Efstathiou, G., 
Frenk, C.~S., White, S.~D.~M., \& Davis, M.\ Gravitational clustering from scale-free initial conditions, \mnras, 235, 715 (1988)
\bibitem{Efstathiou90} Efstathiou, G., 
Sutherland, W.~J., \& Maddox, S.~J.\  
	The cosmological constant and cold dark matter, \nat, 348, 705 (1990)
\bibitem{Einasto1997} Einasto, J., et al., A 120 Mpc Periodicity in the Three-Dimensional Distribution of Galaxy Superclusters, \nat, 385, 139 (1997)
\bibitem{EisensteinHu1998} Eisenstein, D.~J., \& Hu, W.\ Baryonic Features in the Matter Transfer Function, \apj, 496, 605 (1998)
\bibitem{Eisenstein2005} Eisenstein, D.~J., 
et al.\ Detection of the Baryon Acoustic Peak in the Large-Scale Correlation Function of SDSS Luminous Red Galaxies, \apj, 633, 560 (2005)
\bibitem{Ellis2002} Ellis, G.~F.~R., A historical review of how the cosmological constant has fared in general relativity and cosmology, Chaos, Solitons \& Fractals, Elsevier,
46, 505 (2002)
\bibitem{Ellisetal2007} Ellis, G.~F.~R., Maartens, R., \& MacCallum, M.~A.~H., Causality and the speed of sound, General Relativity and Gravitation, 39, 1651 (2007)
\bibitem{Enqvist2008} Enqvist, K., Lema\^\i tre Tolman Bondi model and accelerating expansion, General 
Relativity and Gravitation, 40, 451 (2008)
\bibitem{EnqvistMattsson2007} Enqvist, K., \& Mattsson, T., The effect of inhomogeneous expansion on the supernova observations, Journal of Cosmology and Astro-Particle Physics, 2, 19 (2007)
\bibitem{Kess3} Erickson, J.~K., Caldwell, R.~R., Steinhardt, P.~J., Armendariz-Picon, C., 
\& Mukhanov, V., Measuring the Speed of Sound of Quintessence, Physical Review Letters, 88, 121301 (2002)
\bibitem{Ettori2009}  Ettori, S., Morandi, A., Tozzi, P., Balestra, I., Borgani, S., Rosati, P., Lovisari, L., \& Terenziani, F.\, The cluster gas mass fraction as a cosmological probe: a revised study \aap, 501, 61 (2009)
\bibitem{Evrard1989} Evrard, A.~E.\ Biased cold dark matter theory - Trouble from rich clusters?, \apjl, 341, L71 (1989)
\bibitem{Evrard2008} Evrard, A.~E., et al.\ 
Virial Scaling of Massive Dark Matter Halos: Why Clusters Prefer a High Normalization Cosmology, \apj, 672, 122 (2008)
\bibitem{Fangetal2008} Fang, W., Wang, S., Hu, 
W., Haiman, Z., Hui, L., 
\& May, M., Challenges to the DGP Model from Horizon-Scale Growth and Geometry, ArXiv e-prints, 808, arXiv:0808.2208 (2008)
\bibitem{luisfb} Ferramacho, L. D. \& Blanchard, A. Gas mass fraction from \emph{XMM-Newton} and \emph{Chandra} high redshift clusters and its use as a cosmological test, A\&A, 463, 423 (2007)
\bibitem{lfabyz} Ferramacho, L.~D., 
Blanchard, A., \& Zolnierowski, Y.\ Constraints on CDM cosmology from galaxy power spectrum, CMB and SNIa evolution, \aap, 499, 21 (2009)
\bibitem{HZT2} Filippenko, A.~V., \& Riess, A.~G.\ Results from the High-z Supernova Search Team, Phys.~Rep., 307, 31 (1998)
\bibitem{Fixsen} Fixsen, D.~J., Cheng, 
E.~S., Gales, J.~M., Mather, J.~C., Shafer, R.~A., 
\& Wright, E.~L.\ The Cosmic Microwave Background Spectrum from the Full COBE FIRAS Data Set, \apj, 473, 576 (1996) 
\bibitem{Flanaganetal2000} Flanagan, 
{\'E}.~{\'E}., Tye, S.-H.~H., \& Wasserman, I.\,Cosmological expansion in the Randall-Sundrum brane world scenario, \prd, 62, 0440399
 (2000)
\bibitem{Flanagan2005} Flanagan, {\'E}.~{\'E}.,
Can superhorizon perturbations drive the acceleration of the Universe?, \prd, 71, 103521 (2005)
\bibitem{Freedman2001} Freedman, W.~L., et 
al.\ Final Results from the Hubble Space Telescope Key Project to Measure the Hubble Constant, \apj, 553, 47 (2001)
\bibitem{cardassian} Freese, K., \& Lewis, M., Cardassian expansion: a model in which the universe is flat, matter dominated, and accelerating, Physics Letters B, 540, 1 (2002)
\bibitem{Friedmann1922} Friedmann, A., \"Uber die Kr\"ummung des Raumes , Zs. f\"ur Phys., 10, 377 (1922); english translation in: Friedmann, A., On the curvature of space, General 
Relativity and Gravitation, 31, 1991 (1999)
\bibitem{Friedmann1924} Friedmann, A. , \"Uber die M\"oglichkeit einer Welt mit konstanter negativer Kr\"ommung des Raumes, Zs. f\"ur Phys., 21, 326 (1924); english translation in: Friedmann, A., On the Possibility of a World wih Constant Negative Curvature of Space, General 
Relativity and Gravitation, 31, 2001 (1999)
\bibitem{FournierAlbe} Fournier d'Albe, E.E., Two news Worlds: The infra World; The Supra World, Longmans Green, London (1907)
\bibitem{Futamase1989} Futamase, T.\ An approximation scheme for constructing inhomogeneous universes in general relativity, \mnras, 
237, 187 (1989)
\bibitem{JuanVoid1} Garcia-Bellido, J., \& Haugb{\o}lle, T., Confronting Lemaitre-Tolman-Bondi models with observational cosmology, Journal of Cosmology and Astro-Particle Physics, 4, 3 (2008)
\bibitem{JuanVoid2} Garcia-Bellido, J., \& Haugboelle, T., Looking the void in the eyes - the kSZ effect in LTB models, Journal of Cosmology and Astro-Particle Physics, 9, 16 (2008)
\bibitem{Garfinkle} Garfinkle, D., Inhomogeneous spacetimes as a dark energy model, Classical and Quantum Gravity, 23, 4811 (2006)
\bibitem{Geshnizjani2005} Geshnizjani, G., 
Chung, D.~J., \& Afshordi, N., Do large-scale inhomogeneities explain away dark energy?, \prd, 72, 023517 (2005)
\bibitem{Giovannini2006} Giovannini, M., Inhomogeneous dusty universes and their deceleration, 
Physics Letters B, 634, 1 (2006)
\bibitem{SCP3} Goldhaber, G., The Acceleration of the Expansion of the Universe: A Brief Early History of the Supernova Cosmology Project (SCP),
American Institute of Physics Conference Series, 1166, 53  (2009)
\bibitem{SCP1} Goldhaber, G., \& Perlmutter, S.\ A study of 42 Type Ia supernovae and a resulting measurement of $\Omega_m$ and  $\Omega_\Lambda$ ., Phys. Rep., 307, 325 (1998)
\bibitem{GondoloFreese} Gondolo, P., \& Freese, K., Fluid interpretation of Cardassian expansion, \prd, 68, 063509 (2003)
\bibitem{Grujic2002} Grujic, P.~V., The concept of fractal cosmos: II. Modern cosmology, Serbian 
Astronomical Journal, 165, 45 (2002)
\bibitem{nogo2006} Gruzinov, A., Kleban, 
M., Porrati, M., 
\& Redi, M., Gravitational backreaction of matter inhomogeneities, Journal of Cosmology and Astro-Particle Physics, 12, 1 (2006)
\bibitem{Gush1990} Gush, H.~P., Halpern, M., 
\& Wishnow, E.~H.\ Rocket measurement of the cosmic-background-radiation mm-wave spectrum, Physical Review Letters, 65, 537 (1990)
\bibitem{Guth} Guth, A.~H.\ Inflationary universe: A possible solution to the horizon and flatness problems, \prd, 23, 347 (1981)
\bibitem{hklm} Hamilton, A.~J.~S., 
Kumar, P., Lu, E., \& Matthews, A.\ Reconstructing the primordial spectrum of fluctuations of the universe from the observed nonlinear clustering of galaxies, \apjl, 374, L1 (1991)
\bibitem{H97} Henry, J.~P.\ A Measurement of the Density Parameter Derived from the Evolution of Cluster X-Ray Temperatures, \apjl, 489, 
L1 (1997)
\bibitem{Hill2002} Hill, V., et al.\ First stars. I. The extreme r-element rich, iron-poor halo giant CS 31082-001. Implications for the r-process site(s) and radioactive cosmochronology, \aap, 387, 560 (2002)
\bibitem{HirataSeljak2005} Hirata, C.~M., \& Seljak, U., Can superhorizon cosmological perturbations explain the acceleration of the universe?, \prd, 72, 083501 (2005)
\bibitem{HolzWald} Holz, D.~E., \& Wald, R.~M., New method for determining cumulative gravitational lensing effects in inhomogeneous universes, \prd, 58, 063501 (1998)
\bibitem{HuntSarkar} Hunt, P., \& Sarkar, S., Multiple inflation and the WMAP ``glitches''. II. Data analysis and cosmological parameter extraction, \prd, 76, 123504 (2007)
\bibitem{HuntSarkarVoid} Hunt, P., \& Sarkar, S., Constraints on large scale voids from WMAP-5 and SDSS,  arXiv:0807.4508, \mnras 401, 547 (2010)
\bibitem{Iguchi2002} Iguchi, H., Nakamura, 
T., \& Nakao, K., Is Dark Energy the Only Solution to the Apparent Acceleration of the Present Universe?, Progress of Theoretical Physics, 108, 809 (2002)
\bibitem{Iocco08} Iocco, F., Mangano, G., 
Miele, G., Pisanti, O., 
\& Serpico, P.~D., Primordial Nucleosynthesis: from precision cosmology to fundamental physics, Phys. Rep., 472, 1 (2009)
\bibitem{Ishak2007} Ishak, M., Richardson, 
J., Whittington, D., 
\& Garred, D., Dark Energy or Apparent Acceleration Due to a Relativistic Cosmological Model More Complex than FLRW?, \prd, 78, 123531 (2008)
\bibitem{IshibashiWald2006} Ishibashi, A., \& Wald, R.~M., Can the acceleration of our universe be explained by the effects of inhomogeneities?, Classical and Quantum Gravity, 23, 235 (2006)
\bibitem{Jenkins} Jenkins, A., Frenk, 
C.~S., White, S.~D.~M., Colberg, J.~M., Cole, S., Evrard, A.~E., Couchman, 
H.~M.~P., \& Yoshida, N.\ 
	The mass function of dark matter haloes, \mnras, 321, 372 (2001)
\bibitem{Jetzer&Straumann} Jetzer, P., \& Straumann, N., Has dark energy really been discovered in the Lab?, Physics Letters B, 606, 77 (2005)
\bibitem{Jha2007} Jha, S., Riess, A.~G., 
\& Kirshner, R.~P., Improved Distances to Type Ia Supernovae with Multicolor Light-Curve Shapes: MLCS2k2, \apj, 659, 122 (2007)
\bibitem{BirkhoffR} Johansen, N.~V., \& Ravndal, F., 2006, General Relativity and Gravitation, 38, 537 (2006)
\bibitem{Kaluza}  Kaluza, Th., On the problem of unity in physics, Sitzungsber. Preuss. Akad. Wiss. Phys. Math. Klasse 996 (1921)
\bibitem{NoGo2} Kasai, M., Asada, H., 
\& Futamase, T., Toward a No-Go Theorem for an Accelerating Universe through a Nonlinear Backreaction, Progress of Theoretical Physics, 115, 827 (2006)
\bibitem{KibbleLieu2005} Kibble, T.~W.~B., \& Lieu, R., Average Magnification Effect of Clumping of Matter, \apj, 632, 718 (2005)
\bibitem{Kinney2001}  Kinney, W.~H. \ How to fool cosmic microwave background parameter estimation,  PRD   {63}, 043001 (2001)
\bibitem{Kirshner2002} Kirshner, R.~P, The 
extravagant universe : exploding stars, dark energy and the accelerating 
cosmos / Robert P.~Kirshner.~Princeton, N.J.~: Princeton University Press (2002)
\bibitem{Klein} Quantum theory and five dimensional theory of relativity », Z. Phys. 37 895 (1926)
\bibitem{Koch} Koch, R.~H., van Harlingen, 
D., \& Clarke, J., Quantum-Noise Theory for the Resistively Shunted Josephson Junction, Physical Review Letters, 45, 2132 (1980)
\bibitem{Kolbetal2008} Kolb, E.~W., Marra, V.,  \& Matarrese, S., On the description of our cosmological spacetime as a perturbed conformal 
Newtonian metric and implications for the backreaction proposal for the accelerating universe, ArXiv e-print, \prd, 78, 103002 (2008)
\bibitem{Kolbetal2005} Kolb, E.~W., Matarrese, 
S., Notari, A., \& Riotto, A., Effect of inhomogeneities on the expansion rate of the universe,  \prd, 71, 023524 (2005)
\bibitem{Kolbetal2006} Kolb, E.~W., Matarrese, 
S., \& Riotto, A., On cosmic acceleration without dark energy, New Journal of Physics, 8, 322 (2006)
\bibitem{Komatsu} Komatsu, E., et al.\ 
Five-Year Wilkinson Microwave Anisotropy Probe Observations: Cosmological Interpretation, \apjs, 180, 330 (2009)
\bibitem{Kowalskietal2008} Kowalski, M., et al., Improved Cosmological Constraints from New, Old and Combined Supernova Datasets, \apj, 686, 749 (2008)
\bibitem{KrausChaboyer} Krauss, L.~M., \& Chaboyer, B.\ Age Estimates of Globular Clusters in the Milky Way: Constraints on Cosmology, Science, 299, 65 (2003)
\bibitem{KraussTurner} Krauss, L.~M., \& Turner, M.~S.\  The cosmological constant is back, General Relativity and Gravitation, 27, 1137 (1995)
\bibitem{KumarFlanagan2008} Kumar, N., \& Flanagan, E.~E., Backreaction of superhorizon perturbations in scalar field cosmologies, \prd, 78, 063537 (2008)
\bibitem{ll} Lachi\`eze-Rey, M., \& Luminet, J.\ Cosmic topology, Phys.~Rep., 254, 135 (1995)
\bibitem{Larena09} Larena, J., Alimi, 
J.-M., Buchert, T., Kunz, M., \& Corasaniti, P.-S.\ Testing backreaction effects with observations, \prd, 79, 083011 (2009)
\bibitem{Leithetal2008} Leith, B.~M., Ng, 
S.~C.~C., \& Wiltshire, D.~L., , Gravitational Energy as Dark Energy: Concordance of Cosmological Tests, \apjl, 672, L91 (2008)
\bibitem{Lemaitre1925} Lema{\^i}tre, G., Note on de Sitter's Universe, Journal of Mathematics and Physics 4, 37 (1925)
\bibitem{Lemaitre1927} Lema{\^i}tre, G., Un univers homog\`ene de masse constante et de rayon croissant, rendant compte de la vitesse radiale des n\'ebuleuses extagalactiques,
Annales de la Soci\'et\'e Scientifique de Bruxelles, 47, 49 (1927)
\bibitem{Lemaitre1933} Lema{\^i}tre, G., L'Univers en expansion, 
Annales de la Soci\'et\'e Scientifique de Bruxelles, 53, 51 (1933) (For an english  translation:  General Relativity and Gravitation, 29, 641 (1997)) 
\bibitem{LiSchwarz2007} Li, N., \& Schwarz, D.~J., Signatures of cosmological backreaction, \prd, 78, 083531 (2008)
\bibitem{Lilje} Lilje, P.~B.\ Abundance of rich clusters of galaxies - A test for cosmological parameters, \apjl, 386, 
L33 (1992)
\bibitem{Linden} Linden, S., Virey, J.-M., \& Tilquin, A., Cosmological parameter extraction and biases from type Ia supernova magnitude evolution, \aap, 506, 1095 (2009) 
\bibitem{Linder2005} Linder, E.~V., Cosmic growth history and expansion history, \prd, 72, 
043529 (2005)
\bibitem{Charley1} Lineweaver, C.~H., Barbosa, D., Blanchard, A., \& Bartlett, J.~G.\ Constraints on h, $\Omega_0$ and $\lambda_0$ from cosmic microwave background observations, \aap, 322, 365 (1997)
\bibitem{Charley2} Lineweaver, C.~H., \& Barbosa, D.\ What Can Cosmic Microwave Background Observations Already Say about Cosmological Parameters in Open and Critical-Density Cold Dark Matter Models?, \apj, 496, 624 (1998)
\bibitem{Lindeetal1995} Linde, A., Linde, D., 
\& Mezhlumian, A., Do we live in the center of the world?, Physics Letters B, 345, 203 (1995)
\bibitem{dgp:solarsystem} Lue, A., \& Starkman, G., Gravitational leakage into extra dimensions: Probing dark energy using local gravity, \prd, 67, 064002 (2003)
\bibitem{OP1} Lumb, D.~H.  et al. \  The XMM-NEWTON $\Omega$ project. I. The X-ray luminosity-temperature relation at $z>0.4$, \aap,   {420},  853 (2004) 
\bibitem{luminet} Luminet, J.-P., Weeks, 
J.~R., Riazuelo, A., Lehoucq, R., \& Uzan, J.-P.\ Dodecahedral space topology as an explanation for weak wide-angle temperature correlations in the cosmic microwave background, \nat, 425, 593 (2003)
\bibitem{MaddoxAPM} Maddox, S.~J., 
Efstathiou, G., Sutherland, W.~J., \& Loveday, J.\ Galaxy correlations on large scales, \mnras, 242, 43P (1990)
\bibitem{MadoreFreedman1998} Madore, B.~F., \& Freedman, W.~L., Calibration of the Extragalactic Distance Scale, Stellar astrophysics for the local group: VIII Canary Islands Winter School of Astrophysics, 263 (1998)
\bibitem{MandelbrotCRAS} Mandelbrot, B., Correlations and texture in a new model hierarchical universe, based upon trema sets., 
Academie des Science Paris Comptes Rendus Serie B Sciences Physiques, 288, 
81 (1979)
\bibitem{Mansouri2005} Mansouri, R., Structured FRW universe leads to acceleration: a non-perturbative approach, ArXiv 
Astrophysics e-prints, arXiv:astro-ph/0512605 (2005)
\bibitem{Mansouri2006} Mansouri, R., Illuminating the dark ages of the universe: the exact backreaction in the SFRW model and the acceleration of the universe, ArXiv 
Astrophysics e-prints, arXiv:astro-ph/0601699 (2006) 
\bibitem{Mahajan-et-al2006} Mahajan, G., Sarkar, 
S., \& Padmanabhan, T., Casimir effect confronts cosmological constant, Physics Letters B, 641, 6 (2006)
\bibitem{Firas} Mather, J.~C., et al.,  
A preliminary measurement of the cosmic microwave background spectrum by the Cosmic Background Explorer (COBE) satellite, \apjl, 354, L37 (1990)
\bibitem{Matsubara2004} Matsubara, T.\ Correlation Function in Deep Redshift Space as a Cosmological Probe, \apj, 
615, 573 (2004)
 \bibitem{Mattsson2007} Mattsson, T., Dark energy as a mirage, General 
Relativity and Gravitation, 42, 567  (2010)
 \bibitem{McKellar} McKellar, A., Molecular Lines from the Lowest States of Diatomic Molecules Composed of Atoms Probably Present in Interstellar Space, Publications of the Dominion Astrophysical Observatory (Victoria, BC) 7, 251 ((1941)
\bibitem{Melchiorrietal2003} Melchiorri, A.,
Mersini, L., {\"O}dman, C.~J., \& Trodden, M., The state of the dark energy equation of state, \prd, 68, 043509 (2003)
\bibitem{Michel} Michel, F.~C., Comment on Zero-Point Fluctuations and the Cosmological Constant, \apj, 466, 
660 (1996)
\bibitem{Milgrom} Milgrom, M., A modification of the Newtonian dynamics as a possible alternative to the hidden mass hypothesis, \apj, 270, 
365 (1983)
\bibitem{Moffat2006} Moffat, J.~W., Late-time Inhomogeneity and Acceleration of the Universe, Albert 
Einstein Century International Conference, 861, 987 (2006)
\bibitem{Mohr} Mohr, J.~J., Reese, E.~D., 
Ellingson, E., Lewis, A.~D., \& Evrard, A.~E.\ The X-Ray Size-Temperature Relation for Intermediate-Redshift Galaxy Clusters, \apj, 544, 109 (2000)
\bibitem{Mukhanovetal1997} Mukhanov, V.~F., 
Abramo, L.~R.~W., 
\& Brandenberger, R.~H., Backreaction Problem for Cosmological Perturbations, Physical Review Letters, 78, 1624 (1997)
\bibitem{WFXT} Murray, S.~S., et al.\ Wide field x-ray telescope mission, SPIE, 7011,  46 (2008)
\bibitem{NambuTanimoto} Nambu, Y., \& Tanimoto, M., Accelerating Universe via Spatial Averaging, ArXiv General Relativity and Quantum Cosmology e-prints, arXiv:gr-qc/0507057 (2005)
\bibitem{Jayant2008} Narlikar, J.~V., Burbidge, G., 
\& Vishwakarma, R.~G.,  Cosmology and Cosmogony in a Cyclic Universe, Journal of Astrophysics and Astronomy, 28, 67 (2008)
\bibitem{revDEMG} Nojiri, S., \& Odintsov, S.~D., Introduction to Modified Gravity and Gravitational Alternative for Dark Energy, Int.J.Geom.Meth.Mod.Phys., 4,  115 (2007)
\bibitem{OstrikerSteinhardt} Ostriker, J.~P., \& Steinhardt, P.~J.\  Cosmic Concordance, arXiv:astro-ph/9505066; Ostriker, J.~P., \& Steinhardt, P.~J.\ The observational case for a low-density Universe with a non-zero cosmological constant, \nat, 377, 600 (1995)
\bibitem{OB92} Oukbir, J. \&  Blanchard, A.\ X-ray clusters in open universes,  \aap  {262}, L21 (1992)
\bibitem{OB97} Oukbir, J., \& Blanchard, A.\ X-ray clusters: towards a new determination of the density parameter of the universe, \aap, 317, 1 (1997)
\bibitem{Peacock2008} Peacock, J.~A.\ A diatribe on expanding space, 
arXiv:0809.4573 (2008)
\bibitem{Peebles1980} Peebles, P.~J.~E.\   
 The large-scale structure of the universe, 
Princeton University Press, Princeton, N.J., (1980) 
\bibitem{PeeblesCDM} Peebles, P.~J.~E.\ Large-scale background temperature and mass fluctuations due to scale-invariant primeval perturbations, 
\apjl, 263, L1 (1982)
\bibitem{Peebles84} Peebles, P.~J.~E.\  Tests of cosmological models constrained by inflation, \apj, 284, 439 (1984)
\bibitem{Peebles89} Peebles, P.~J.~E., 
Daly, R.~A., \& Juszkiewicz, R.\ Masses of rich clusters of galaxies as a test of the biased cold dark matter theory, \apj, 347, 563 (1989)
\bibitem{Peebles93} Peebles, P. J. E., Principles of physical cosmology, Princeton Series in Physics, Princeton, NJ: Princeton University Press (1993)
\bibitem{PeeblesRatra1988} Peebles, P.~J.~E., \& Ratra, B., Cosmology with a time-variable cosmological 'constant', \apjl, 325, L17 (1988)
\bibitem{Percival01} Percival, W.~J., et 
al., The 2dF Galaxy Redshift Survey: the power spectrum and the matter content of the Universe.   \mnras, 327, 1297 (2001)
\bibitem{Percivaletal2007} Percival, W.~J., et 
al.\ The Shape of the Sloan Digital Sky Survey Data Release 5 Galaxy Power Spectrum, \apj, 657, 645 (2007)
\bibitem{CasimirLambda} Perivolaropoulos, L., Vacuum energy, the cosmological constant, and compact extra dimensions: Constraints from Casimir effect experiments, \prd, 77, 107301 (2008)
\bibitem{SCP2} Perlmutter, S., et 
al.\ Measurements of Omega and Lambda from 42 High-Redshift Supernovae, \apj, 517, 565 (1999)
\bibitem{Perrenod} Perrenod, S.~C.\ The evolution of cluster X-ray sources. IV - The luminosity function, \apj, 236, 373 (1980)
\bibitem{Pierpaoli} Pierpaoli, E., 
Borgani, S., Scott, D., \& White, M.\ On determining the cluster abundance normalization, \mnras, 342, 163 (2003)
\bibitem{PS} Press, W.~H., \& Schechter, P.\ Formation of Galaxies and Clusters of Galaxies by Self-Similar Gravitational Condensation, \apj, 187, 425 (1974)
\bibitem{Rasanen2006} R{\"a}s{\"a}nen, S., Constraints on backreaction in dust universes, Classical and Quantum Gravity, 23, 1823 (2006)
\bibitem{Rasanen2008} Rasanen, S., The effect of structure formation on the expansion of the universe, ArXiv 
e-prints, 805, arXiv:0805.2670 (2008)
\bibitem{RatraPeebles1988} Ratra, B., \& Peebles, P.~J.~E., Cosmological consequences of a rolling homogeneous scalar field, \prd, 37, 3406 (1988)
\bibitem{acbar2008} Reichardt, C.~L., et 
al.\ High resolution CMB power spectrum from the complete ACBAR data set, arXiv:0801.1491 (2008)
\bibitem{R99} Reichart, D.~E., 
Nichol, R.~C., Castander, F.~J., Burke, D.~J., Romer, A.~K., Holden, B.~P., 
Collins, C.~A., \& Ulmer, M.~P.\ A Deficit of High-Redshift, High-Luminosity X-Ray Clusters: Evidence for a High Value of $\Omega_m$?, \apj, 518, 521 (1999)
\bibitem{Reiprich2006} Reiprich, T.~H.\ The galaxy cluster X-ray luminosity-gravitational mass relation in the light of the WMAP 3rd year data, \aap, 453, L39 (2006)
\bibitem{Riess98} Riess, A.~G., et al., Observational Evidence from Supernovae for an Accelerating Universe and a Cosmological Constant
 \aj, 116, 1009 (1998)
\bibitem{Rines2007} Rines, K., Diaferio, A., 
\& Natarajan, P.\ The Virial Mass Function of Nearby SDSS Galaxy Clusters, \apj, 657, 183 (2007)
\bibitem{Rubinetal2008} Rubin, D., et al., Looking Beyond Lambda with the Union Supernova Compilation,  
\apj, 695, 391 (2009)
\bibitem{dgp:obs} Rydbeck, S., Fairbairn,  M., 
\& Goobar, A., Testing the DGP model with ESSENCE, Journal of Cosmology and Astro-Particle Physics, 5, 3 (2007)
\bibitem{SachsWolfe} Sachs, R.~K., \& Wolfe, A.~M., Perturbations of a Cosmological Model and Angular Variations of the Microwave Background, \apj, 147, 73  (1967)
\bibitem{S97} Sadat, R., Blanchard, A., \& Oukbir, J.\ Constraining $\Omega_0$ from X-ray properties of clusters of galaxies at high redshift, \aap, 329, 21 (1998)
\bibitem{sadat05} Sadat, R., Blanchard, A., Vauclair, S. C. et al., New light on the baryon fraction in galaxy clusters  , {A\&A}, 437, 310 (2005)
\bibitem{Sasaki} Sasaki, S.\ A New Method to Estimate Cosmological Parameters Using the Baryon Fraction of Clusters of Galaxies, \pasj, 48, 
L119 (1996)
\bibitem{SanchezCole} S{\'a}nchez, A.~G., \& Cole, S.\ The galaxy power spectrum: precision cosmology from large-scale structure?, \mnras, 385, 830 (2008)
\bibitem{Sanchez} S{\'a}nchez, A.~G., Crocce, 
M., Cabre, A., Baugh, C.~M., \& Gaztanaga, E.\ Cosmological parameter constraints from SDSS luminous red galaxies: a new treatment of large-scale clustering, \mnras, 400, 1643 (2009)
\bibitem{SwissCheese} Sch{\"u}cking, E., Das Schwarzschildsche Linienelement und die Expansion des Weltalls.,  Zeitschrift fur Physik , 137, 595 (1954)
\bibitem{ST} Sheth, R.~K., \& Tormen, G.\ Large-scale bias and the peak background split, \mnras, 308, 119 (1999)
\bibitem{ST2} Sheth, R.~K.,  Mo, H.~J. \& Tormen, G.\ Ellipsoidal collapse and an improved model for the number and spatial distribution of dark matter haloes, MNRAS  {323}, 1 (2001)
\bibitem{teves:obs} Skordis, C., Mota, 
D.~F., Ferreira, P.~G., 
\& B{\oe}hm, C., Large Scale Structure in Bekenstein's Theory of Relativistic Modified Newtonian Dynamics, Physical Review Letters, 96, 011301 (2006)
\bibitem{1992ApJ...396L...1S} Smoot, G.~F., et al.\ 
Structure in the COBE differential microwave radiometer first-year maps, \apjl, 396, L1 (1992)
\bibitem{Songaila1994} Songaila, A., et al.\ Measurement of the Microwave Background Temperature at a Redshift of 1.776, \nat, 371, 43 (1994)
\bibitem{spergel2003} Spergel, D.~N., et al.\ 
First-Year Wilkinson Microwave Anisotropy Probe (WMAP) Observations: Determination of Cosmological Parameters, \apjs, 148, 175 (2003)
\bibitem{Srianand2008} Srianand, R., Noterdaeme, P., Ledoux, C., \& Petitjean, P.\ First detection of CO in a high-redshift damped Lyman-$\alpha$ system, \aap, 482, L39 (2008)
\bibitem{Tammann2008} Tammann, G.~A., Sandage, A., \& Reindl, B., The expansion field: the value of H$_0$,  \aapr, 15, 289 (2008)
\bibitem{Tegmarketal2004} Tegmark, M., et al.\ 
The Three-Dimensional Power Spectrum of Galaxies from the Sloan Digital Sky Survey, \apj, 606, 702 (2004)
\bibitem{Tegmarketal2006} Tegmark, M., et al.\ Cosmological constraints from the SDSS luminous red galaxies, \prd, 74, 123507 (2006)
\bibitem{Tinker2008} Tinker, J., Kravtsov, 
A.~V., Klypin, A., Abazajian, K., Warren, M., Yepes, G., Gottl{\"o}ber, S., 
\& Holz, D.~E.\ Toward a Halo Mass Function for Precision Cosmology: The Limits of Universality, \apj, 688, 709 (2008)
\bibitem{Tolman1934} Tolman, R.~C., Effect of Inhomogeneity on Cosmological Models, Proceedings of the National Academy of Science, 20, 169 (1934)
\bibitem{Tomita2001} Tomita, K., A local void and the accelerating Universe, \mnras, 326, 287 (2001)
\bibitem{Tomita2001b} Tomita, K., Analyses of Type Ia Supernova Data in Cosmological Models with a Local Void, Progress of 
Theoretical Physics, 106, 929 (2001)
\bibitem{earlyBR} Tsamis, N.~C., \& Woodard, R.~P., Relaxing the cosmological constant., Physics Letters B, 301, 351 (1993)
\bibitem{Uzan} Uzan, J.-P., The acceleration of the universe and the physics behind it, Gen. Rel. Grav. 39, 307 (2007)
\bibitem{Uzan2008} Uzan, J.-P., Clarkson, C., 
\& Ellis, G.~F.~R., Time Drift of Cosmological Redshifts as a Test of the Copernican Principle, Physical Review Letters, 100, 191303 (2008)
\bibitem{Vanderveldetal2006} Vanderveld, R.~A., 
Flanagan, {\'E}.~{\'E}., \& Wasserman, I., Mimicking dark energy with Lema\^\i tre-Tolman-Bondi models: Weak central singularities and critical points, \prd, 74, 023506 (2006)
\bibitem{Vauclair03} Vauclair, S.~C., et al.\ The XMM-Omega project. II. Cosmological implications from the high redshift L - T relation of X-ray clusters, \aap, 412, L37 (2003)
\bibitem{VL} Viana, P.~T.~P., \& Liddle, A.~R.\ Galaxy clusters at $0.3<z<0.4$ and the value of $\Omega_0$, \mnras, 303, 535 (1999)
\bibitem{coldspot} Vielva, P., 
Mart{\'{\i}}nez-Gonz{\'a}lez, E., Barreiro, R.~B., Sanz, J.~L., 
\& Cay{\'o}n, L.,  Detection of Non-Gaussianity in the Wilkinson Microwave Anisotropy Probe First-Year Data Using Spherical Wavelets, \apj, 609, 22 (2004)
\bibitem{V02} Vikhlinin, A.  et al. \ Evolution of the Cluster X-Ray Scaling Relations since z $>$ 0.4,  ApJL  {578}, 107 (2002)
\bibitem{Vik09} Vikhlinin, A., et 
al.\ Chandra Cluster Cosmology Project. II. Samples and X-Ray Data Reduction, \apj, 692, 1033 (2009) 
\bibitem{Warren} Warren, M.~S., 
Abazajian, K., Holz, D.~E., \& Teodoro, L.\ Precision Determination of the Mass Function of Dark Matter Halos, \apj, 646, 881 ( 2006)
\bibitem{WeinbergI} Weinberg, S.,\  Gravitation and Cosmology: Principles and Applications of the General Theory of Relativity,pp.~688.~ISBN 0-471-92567-5.~Wiley-VCH  (1972)  
\bibitem{Weinberg76} Weinberg, S.,  Apparent luminosities in a locally inhomogeneous universe, \apjl, 208, L1 (1976)
\bibitem{Weinberg2008} Weinberg, S.,\  
Cosmology,~ISBN 978-0-19-852682-7.~Published by Oxford 
University Press, Oxford, UK (2008)
\bibitem{Wetterich2003} Wetterich, C.,  Can structure formation influence the cosmological evolution?, \prd, 
67, 043513 (2003)
\bibitem{White1993}
White, S.D.M., Navarro, J.F., Evrard, A.E. \& Frenk, C., The baryon content of galaxy clusters: a challenge to cosmological orthodoxy, Nature 366, 429 (1993)
\bibitem{White07} White, S.~D.~M.\  Fundamentalist physics: why Dark Energy is bad for astronomy, Reports 
on Progress in Physics, 70, 883 (2007)
\bibitem{Wiltshire2007} Wiltshire, D.~L., Cosmic clocks, cosmic variance and cosmic averages, New Journal of Physics, 9, 377 (2007)
\bibitem{Wirtz1} Wirtz, C. Einiges zur Statistik des Radialgeschhwindigkeiten von Spiralnebeln und Kugelsternhaufen, Astr. Nachr., 222, 21 (1922)
\bibitem{Wirtz2} Wirtz, C.,  De Sitters Kosmologie und die Radialbewegungen des Spiralnebel, Astr. Nachr., 222, 21 (1924); see also Seitter C. \& 
Duerbeck, W. 1990, in ``Modern Cosmology in Retrospect'', Ed. by B.Bertotti, R. Balbinot S. Bergia A.Messina, Cambridge University press,  365
\bibitem{wrightSN} Wright, E.~L.\ Distant Supernovae and the Accelerating Universe, 
arXiv:astro-ph/0201196 (2002)
\bibitem{NedCMB} Wright, E.~L. for a summary of pre-discovery of the CMB
\url{http://www.astro.ucla.edu/~wright/CMB.html}
\bibitem{COBEpar} Wright, E.~L., et al.\ Interpretation of the cosmic microwave background radiation anisotropy detected by the COBE Differential Microwave Radiometer, \apjl, 396, L13 (1992)
\bibitem{Zibin2008} Zibin, J.~P., Scalar Perturbations on Lema\^\i tre-Tolman-Bondi spacetimes, \prd, 78, 043504  (2008)

\end{thebibliography}


\end{document}